\documentclass[final,5p,times,twocolumn]{elsarticle}

\usepackage{amssymb}
\usepackage{amsmath}
\usepackage{lineno}
\newcommand{\sech}{\mbox{ sech}}

\usepackage{hyperref}
\hypersetup{
    colorlinks=true,
    linkcolor=black,
    filecolor=magenta,      
    urlcolor=black,
    pdftitle={Overleaf Example},
    pdfpagemode=FullScreen,
    }

\journal{Physcics Letter A}

\begin{document}

\begin{frontmatter}

\title{Propagation of Gaussian beam in strongly nonlocal nonlinear media with inhomogeneous diffraction} 

\author[1]{K. L. Mamatha} 
\author[2]{S. Stalin} 
\ead{stalin.cnld@gmail.com}
\author[1]{K. Nithyanandan*}
\ead{nithyan@phy.iith.ac.in}

\affiliation[1]{organization={Ultrafast Fiber Optics \& Smart Photonic Technologies Lab, Department of Physics, IIT Hyderabad},
            addressline={Kandi}, 
            city={Hyderabad},
            postcode={502282}, 
            state={Telangana},
            country={India}}

 \affiliation[2]{organization={Department of Nonlinear Dynamics, Bharathidasan University},
            addressline={Tiruchirappalli}, 
            postcode={620024}, 
            state={Tamil Nadu},
            country={India}}

\begin{abstract}
Solitons in a strongly nonlocal nonlinear medium have attracted considerable attention in recent years due to their unique and distinctive characteristics from those observed in the local nonlinear medium. In this work, we investigate the propagation of a Gaussian beam in strongly nonlocal nonlinear media with spatially varying diffraction.  Using analytical and numerical approaches, we examine the propagation dynamics and stability characteristics of the nonlocal soliton. For representative cases of diffraction profiles, our results show that diffraction tailoring strongly influences key beam properties, including amplitude, beam width, phase-front curvature, and phase-space evolution. All diffraction modulation supports the formation of diffraction-managed breather solitons, with excellent agreement between theory and simulations.  Furthermore, the modulation instability (MI) studies systematically explored the stability characteristics under various diffraction landscapes. The obtained results demonstrate that by engineering diffraction in such nonlocal media, one can effectively control beam dynamics, thereby opening new possibilities for applications in nonlinear light control, beam shaping, and all-optical system design.
\end{abstract}


\end{frontmatter}

\section{Introduction}
\label{Introduction}

    Solitons are localized self-confined wave packets, that retain their shape during propagation through a medium, arising from a delicate balance between dispersive and nonlinear effects. They have been widely observed in various fields of science, including hydrodynamics, plasma physics, superconductivity, matter waves, and optics, to mention a few~\cite{scott2003nonlinear}. Based on the nature of the balancing effects and the propagation medium, the soliton can be called a temporal or spatial soliton. The former results from the balance of the dispersion with nonlinearity, while the latter emerges from a balance between diffraction and nonlinearity~\cite{kivshar2003optical}. Over the years, the propagation of optical solitons has been extensively investigated in Kerr media, where the change in refractive index is a function of local intensity, governed by the local nonlinear Schr\"odinger equation (NLSE) \cite{stegeman1999optical}.

    Recent advances in nonlinear optics reveal that the refractive index of a material at a given point can depend on the beam intensity at neighbouring points. Such media are referred to as nonlocal nonlinear media, in which the propagation of optical beams or pulses is governed by the nonlocal nonlinear Schr\"odinger equation (NNLSE) \cite{snyder1997accessible}. The degree of nonlocality in such media is characterized by the relationship between the width of the material's response function ($L_c$) and the width of the optical beam ($\omega $) \cite{krolikowski2001modulational,rasmussen2005theory, krolikowski2000solitons, rosanov2002internal, guo2004large, snyder1997accessible}. Based on their relation, nonlocality is broadly categorized into three types: (i) weakly nonlocal ($\omega >L_c$): the width of the optical beam is larger than the length of the response function, (ii) strongly nonlocal ($\omega <L_c$): the optical beam has a narrower width compared to the length of the response function, and (iii) generally nonlocal: the width of the optical beam is comparable with the length of the response function.  Each of these three types of nonlocal nonlinear regimes supports its own class of solitons. For example, weakly nonlocal spatial solitons were reported in Ref. \cite{krolikowski2000solitons}, strongly nonlocal spatial accessible solitons (ASs) were identified in Ref. \cite{snyder1997accessible, zhong2016spatiotemporal, he2008crescent}, and generally nonlocal spatial solitons were discussed in Ref. \cite{ouyang2006perturbative}.

    These nonlocal nonlinearities often result from a transport process, including heat conduction \cite{10.1063/1.1713919,1074930}, molecular or atomic diffusion \cite{PhysRevA.48.4583}, and the diffusion of electric charges in photorefractive media \cite{segev1992photorefractive,Gatz:98}. The nonlocal nature can also be induced by a long-range molecular interaction in dipolar Bose-Einstein condensates (BECs) \cite{PhysRevLett.94.160401} and in nematic liquid crystals (NLCs) \cite{MCLAUGHLIN199555, PhysRevE.66.036603, conti2003route, conti2004observation}. It is also important to note that the highly nonlocal nonlinear nature can be observed in plasmas \cite{Hlpecseli1980}.  Thus, nonlocality is a generic feature of many physical systems, so it is essential to investigate the impact of nonlocal nonlinearity on beam dynamics.  

    In the context of nonlinear optics, various theoretical and numerical studies have investigated soliton dynamics under different strengths of nonlocality by considering constant diffraction. In Ref. \cite{snyder1997accessible}, Snyder and Mitchell investigated the propagation of optical beams in strongly nonlocal nonlinear media by using an assumption is that refractive index change is determined by the total power of the optical beam rather than its local intensity. Then, they simplified the NNLSE to a linear harmonic oscillator equation to understand the dynamics of solitons.  After their work, solitons in strongly nonlocal nonlinear media have attracted significant interest from both theoretical and experimental perspectives \cite{rasmussen2005theory, krolikowski2001modulational, rosanov2002internal, krolikowski2000solitons, ouyang2006perturbative, guo2004large, mihalache2006three,he2008spinning, zhou2014propagation} due to their unique characteristics.  For instance, they exhibit large phase shifts \cite{guo2004large}, attractive interactions between out-of-phase solitons \cite{snyder1997accessible, rasmussen2005theory, peccianti2002nonlocal, hu2006nonlocality,chen2015interaction}, attraction dynamics of the dark solitons \cite{dreischuh2006observation}, and long-range interactions between solitons~\cite{rotschild2006long}. These phenomena are markedly distinct from those observed in local nonlinear media. Additionally, solitons in strongly nonlocal systems undergo periodic evolution during propagation \cite{buccoliero2009quasi, dai2015propagation}, highlighting the profound impact of nonlocality on their dynamics.  Despite the above developments in the strongly nonlocal case, there have also been studies related to other types of nonlocality, such as weakly nonlocal and generally nonlocal media \cite{PhysRevE.63.016610,PhysRevA.82.063829,rasmussen2005theory,kong2010analytical,Tur85,PhysRevE.74.036622} .Considering the unique features of solitons in strongly nonlocal media, we restrict to the strongly nonlocal nonlinear regime in the present work. We note that spatial solitons have been extensively investigated in the context of NLCs \cite{assanto2020spin, sala2017bending, rasmussen2005theory, huang2020diffraction} and they were experimentally observed in planar NLCs \cite{peccianti2000electrically, conti2004observation, peccianti2004routing, peccianti2005interplay, assanto2003nematicons}.
 
    Most of the existing studies on soliton dynamics in nonlocal nonlinear media have been carried out within the framework of homogeneous systems, where the diffraction coefficient remains constant along the propagation direction. An interesting question that arises here is: what will be the effect on the beam characteristics if the medium supports variable diffraction? In the context of spatial solitons, diffraction management plays a role similar to dispersion management in temporal solitons, providing valuable insights into the behavior of light waves in an array of waveguides. The diffraction management is local in nature, so it is of interest to investigate the interplay between local and nonlocal effects in order to better understand their combined influence on beam propagation in a strongly nonlocal nonlinear medium.

    In optical fibers and waveguide arrays, the management of diffraction or dispersion is influenced by inherent inhomogeneities arising from geometric and material imperfections. Geometric irregularities, such as variations in fiber diameter caused by fabrication defects or environmental factors, distort guiding properties and affect light propagation. Material inhomogeneities, including local variations in lattice structure or molecular arrangement, alter the refractive index and consequently influence dispersion, diffraction, phase modulation, and optical gain or loss. In NLCs, inhomogeneities can be introduced by modifying the spatial distribution of molecular orientation through suitable anchoring conditions \cite{huang2020diffraction, sala2017bending, calisto2023direction}. Since dipolar BECs can exhibit strong nonlocal nonlinearity, one can also investigate the effect of inhomogeneity on the soliton structure by tailoring temporal modulations. This can be achieved by tuning the $s$-wave contact interactions through a magnetically or optically induced Feshbach resonance mechanism \cite{sabari2015dynamical,cuevas2009solitons}.

   Understanding nonlinear wave propagation in inhomogeneous nonlocal media is therefore crucial, not only for the advancement of theoretical models but also for improving the design and performance of optical systems used in telecommunications, sensing, and optical switching etc...

    From earlier studies, it is evident that much less effort has been taken for understanding the soliton dynamics in  inhomogeneous strongly nonlocal nonlinear media, where the diffraction varies along the propagation distance. For instance, a systematic investigation in this direction has been done by Mishra \textit{et al.} in \cite{mishra2022energy}, where they have examined the stability and provided energy budget for diffraction-managed breathing accessible solitons  in a strongly nonlocal nonlinear medium. In their work, they considered a periodic arrangement of alternating positive and negative diffraction segments in a discrete fashion. However, to the best of our knowledge, continuously varying specific forms of diffraction along the propagation direction have not been reported anywhere in the literature. In particular, the effects of linear and exponentially decaying diffraction modulations, as well as step-like ($\tanh$), barrier-type ($ \text{sech} $), and periodic ($\sin$) diffraction modulations, on beam dynamics have not yet been unraveled. In the local case, these five different diffraction modulations are strongly correlated with experiments, therefore, it is natural to examine their effects in the nonlocal case. Motivated by this gap in the literature, the present work focuses on the deformations in the propagation characteristics of nonlocal solitons arising from five distinct types of diffraction modulations together with the nonlocal effect.
  
    For this purpose, we consider the evolution of optical Gaussian beams in an inhomogeneous, strongly nonlocal nonlinear medium (SNNM), modeled by a variable-coefficient NNLSE (vc-NNLSE) with five distinct forms of diffraction modulations that widely used in the local or conventional Kerr medium. To address this problem, one may consider the Snyder–Mitchell approximation \cite{snyder1997accessible} or solve the vc-NNLSE directly with a Gaussian response function. In the present work, we adopt the latter approach and assume a Gaussian beam ansatz as the initial condition. Subsequently, the Lagrangian variational method (LVM) is applied to obtain a set of coupled ordinary differential equations (ODEs) governing the beam characteristics, derived from the corresponding averaged Lagrangian density. Further, using the fourth-order Runge-Kutta (RK4) method, we analyzed these system of ODEs numerically. To validate the results, direct simulations of the vc-NNLSE are carried out using the split-step Fourier transform (SSFT) technique. 

    The rest of the paper is structured as follows. Section 2 introduces the vc-NNLS model to describe the optical beam evolution in a SNNM with non-uniform diffraction. In this section, we also describe the Lagrangian variational method relevant to the present work and derive a system of coupled ODEs that govern the evolution of the beam parameters. Section 3 presents the results associated with the numerical investigations of the beam dynamics and it also presents an analysis of the influence of the diffraction modulation effects on the beam characteristics. In Section 4, we examine the modulational instability (MI) of a plane wave in an inhomogeneous SNNM. Finally, Section 6 summarizes the outcomes of the present study and provides appropriate conclusions. In the Appendix, we indicate how to obtain the NNLSE with variable diffraction.

\section{Model and Variational Approach}
  
    \label{Model}
    The propagation of an optical beam in the strongly nonlocal nonlinear medium with variable diffraction is described by the following dimensionless vc-NNLSE, which is of the form,

    \begin{equation}
         i \frac{\partial \psi}{\partial z}+\frac{\mu(z)}{2} \frac{\partial^2 \psi}{\partial x^2}+\rho \psi \int_{-\infty}^{\infty} R(x-\xi) I(\xi, z) d \xi=0.
          \label{NNLSE}
    \end{equation}
    Here,  $\psi(x,z)$ is the slowly varying envelope, $\rho = k \eta$ denotes the nonlinearity coefficient (in this work, we consider focusing nonlinearity $\rho>0$), where $k$ is the wave number, and $\eta$ is a constant that depends on the material properties,  and  $\mu(z)$ is the diffraction coefficient that varies along the propagation direction. In NLCs, $x$ and $z$ are the transverse and longitudinal coordinates, respectively, whereas in dipolar BECs these independent variables correspond to the spatial and temporal coordinates. Variable diffraction $\mu(z)$, or more generally inhomogeneity, can be realized in NLCs by engineering the optical axis along the propagation direction. This can be achieved by modifying the spatial distribution of molecular orientation through suitable anchoring mechanisms. In dipolar BECs, temporal modulation can be implemented by tuning the $s$-wave atomic scattering length via a Feshbach resonance mechanism. The second term accounts for the diffraction of the beam. The convolution integral appearing in the third term constitutes the nonlocal nonlinearity. As shown in \cite{rasmussen2005theory}, the integrable form of the nonlocal nonlinearity appearing in Eq. (\ref{NNLSE}) can be realized in a single transverse-dimensional NLC by expressing the orientation angle of the liquid crystal molecules in terms of the optical field intensity. Equation (\ref{NNLSE}) also arises in dipolar BECs, as a quasi-one-dimensional nonlocal Gross-Pitaevskii equation, where the nonlocal nonlinearity appears as a result of the long-range and anisotropic dipole-dipole interactions.
    
    The NNLSE Eq. (\ref{NNLSE}) can be rewritten as the following linear Schrödinger-like equation in quantum mechanics, that is  
    \begin{equation}
           i \frac{\partial \psi}{\partial z}+\frac{\mu(z)}{2} \frac{\partial^2 \psi}{\partial x^2}+ V(x,z) \psi =0.
          \label{}
    \end{equation}
    In the above equation, $V(x,z)$ is the beam-induced "potential" at a given spatial point $x$.~(\textit{ i.e.,})
    \begin{equation}
          V(x,z) =\rho \int_{-\infty}^{\infty} R(x-\xi) I(\xi, z) d \xi,
    \end{equation}
    where, $ I(\xi, z)$ is the intensity of the paraxial beam, and $R(x)$ is the nonlocal response function. In the present work, we are considering the Gaussian nonlocal response function, which is given by
    \begin{equation}
    R(x)=\frac{1}{\sqrt{\pi} \sigma} \exp \left(-\frac{x^2}{\sigma^2}\right),
    \end{equation}
    where $\sigma=L_c$ is the length of the response function and the kernel ($R(x)$) must satisfy the condition $ \int R(x) dx =1$. Note that one can also use other forms of response functions, such as the exponential-decay response function \cite{krolikowski2001modulational, jia2012solitons}, the rectangular response function \cite{krolikowski2001modulational, kong2013dark}, and the trigonometric oscillation response function \cite{liang2023solitons, Zhang:22}, which have been used in the literature.

\begin{figure*}[hbt!]
         \centering
         \includegraphics[scale = 0.5]{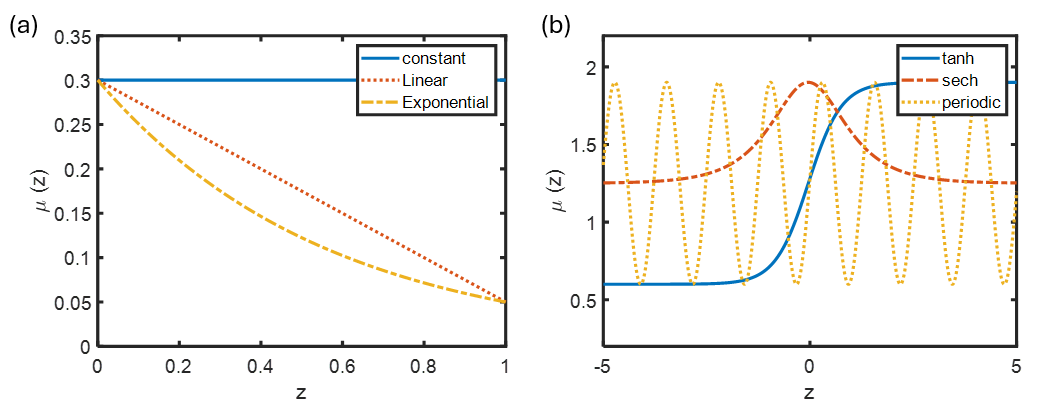}
         \caption{Schematic of the six distinct diffraction profiles, including the constant diffraction case. Panel (a) displays the constant, linearly decreasing, and exponentially decreasing diffraction profiles, while Panel (b) shows the step-like ($\tanh$), barrier-type ($\sech$), and periodic ($\sin$) diffraction profiles.}
         \label{diffraction profile}
    \end{figure*}

    In view of the contemporary interest in conventional diffraction-managed spatial solitons supported by different classes of waveguides, we consider a variety of diffraction-managed profiles in order to investigate Gaussian beam propagation in SNNM. These profiles are as follows ~\cite{da1999analysis,xiong2022analytical, malomed1997pulse}
\\
    Linear diffraction profile :
    \begin{equation}
         \mu(z) = \mu_0 \left[\frac{1-\beta}{\beta L}z +  1\right], \label{linear-diffraction}
    \end{equation}
    Exponentially decaying diffraction profile :
    \begin{equation}
        \mu(z) = \mu_0 \exp{\left[- \frac{z}{L} \log(\beta))\right]}
         \label{exponential-diffraction}
    \end{equation}
    Step-like ($\tanh$) diffraction profile :
    \begin{equation}
            \mu(z) = \mu_0 +\mu_1  \tanh(\mu_2 z + \mu_3)
          \label{tanh}
    \end{equation}
    Barrier-type ($\sech$) diffraction profile :
    \begin{equation}
       \mu(z) = \mu_0 +\mu_1  \sech(\mu_2 z + \mu_3)
       \label{sech}
    \end{equation}
    Periodic profile :
    \begin{equation}
        \mu(z) = \mu_0 + \mu_1  \sin(\mu_2 z + \mu_3).
        \label{periodic}
    \end{equation}
    In the above, $L$ denotes the length of the medium, $1/\beta$ is the minimum value of the $\mu(z)$ in the case of the linear and exponentially decreasing diffraction case, $\mu_0$ is the constant value of the diffraction when there is no inhomogeneity and $\mu_0$, $\mu_1$, $\mu_2$, $\mu_3$ are arbitrary real constants. These five distinct forms of longitudinally varying diffraction modulations, along with the constant case, are illustrated in Figs. \ref{diffraction profile}(a) and \ref{diffraction profile}(b).

    \label{VariationalApproach}
    We investigate the propagation dynamics of the optical beam in an inhomogeneous SNNM by using the LVM
    \cite{anderson1983variational}. This approximate yet powerful analytical approach allows us to reduce the governing vc-NNLSE (\ref{NNLSE}) into a set of coupled ordinary differential equations, which effectively capture the essential beam characteristics and make the problem more tractable. To apply the LVM to Eq. (\ref{NNLSE}), we consider the corresponding Lagrangian density of the form
    \begin{multline}
        L=\frac{i}{2}\left(\psi^* \frac{\partial \psi}{\partial z}-\psi \frac{\partial         \psi^*}{\partial z}\right)-\frac{\mu(z)}{2}\left|\frac{\partial \psi}{\partial  x}\right|^2 \\+
        \frac{\rho}{2} |\psi|^2\int_{-\infty}^{\infty} R(x-\xi) I(\xi, z) d \xi.
       \label{Lagrangian density}
    \end{multline}
    The solution of Eq.~(\ref{NNLSE}) is considered in the form of a Gaussian ansatz 
    \begin{multline}
       \psi(x, z)=A(z) \exp \left(-\frac{x^2}{2 \omega^2(z)}\right)         \exp \left(i c(z) x^2+i \theta(z)\right),
        \label{ansatz}
    \end{multline} 
    where $A(z)$ and $\omega(z)$ represent the amplitude and beam width, respectively.  $c(z)$ is the phase front curvature and $\theta(z)$ is the beam phase.\\ 
 
    The average Lagrangian of the system is found by using the standard Rayleigh-Ritz optimization procedure \cite{anderson1999pereira}, that is, $ \langle L\rangle = \int_{-\infty}^{\infty} L\,dx $. \,By substituting the ansatz given in Eq.~(\ref{ansatz}) into the lagrangian density of the system given in Eq.~(\ref{Lagrangian density}),  we get the average Lagrangian of the system as follows,
    \begin{multline}
         \langle L\rangle  = \int_{-\infty}^{\infty} L \mathrm{~d} x
        =  \frac{\sqrt{\pi} \rho \omega^2 A^4}{2 \sqrt{2 w^2+\sigma^2}} \\
         -\sqrt{\pi} A^2\left( \omega \frac{\partial \theta}{\partial z}+\frac{\omega ^3}{2} \frac{\partial c}{\partial z}+\mu{(z)} \frac{\left(1+4 c^2 \omega^4\right)}{4 \omega}\right).
    \end{multline}
    The average Lagrangian of the system is now varied with respect to the different beam parameters $r_j = A(z), ~c(z),~\omega(z),~ \text{and} ~\theta(z)$ according to the Euler-Lagrangian equation:
    \begin{equation}
         \frac{\partial\langle L\rangle}{\partial r_j}-\frac{\partial}{\partial z}\left(\frac{\partial\langle L\rangle}{\partial\left(\frac{\partial r_j}{\partial z}\right)}\right)=0.
    \end{equation}
    This procedure gives rise to a set of coupled ODEs corresponding to the different beam parameters as follows:
    
    \begin{subequations}
      \begin{eqnarray}
            &&\frac{\partial A}{\partial z} = - \mu A ~ c, \label{ampitude}    \\
            &&\frac{\partial \omega}{\partial z}=2 c\omega  \mu,\label{beamwidth}\\
            &&\frac{\partial c}{\partial z}=\mu\left(-2 c^2+\frac{1}{2 \omega^4}\right)-\frac{ \rho P_0}{ \sqrt{\pi}\left(\sigma^2 + 2 \omega^2\right)^{3 / 2}},\label{curvature}\\
            &&   \frac{\partial \theta}{\partial z}=\frac{\rho P_0\left(5 \omega^2 + 2 \sigma^2\right)}{2 \sqrt{\pi}\left(\sigma^2 +  2 \omega^2\right)^{3 / 2}} - \frac{\mu}{2 \omega^2}.\label{phase}
        \end{eqnarray}
    \end{subequations}
    
    The initial beam energy $P_0$ is found as $P_0 = \sqrt{\pi} \omega A^2$ from  the equation $P_0=\int_{\infty}^{+\infty}|\psi|^2dx$. Further, the critical power ($P_c$) of the system for the soliton propagation in the absence of the diffraction management is determined by using the Eq.~(\ref{curvature}) at the steady state. By setting the phase-front curvature as constant. i.e., $\frac{\partial c}{\partial z} =0$, we obtain the analytical expression for the critical power as
    \begin{equation}
         P_c = \frac{\sqrt{\pi} \mu (\sigma^2 + 2 \omega^2)^{3/2} (1- 4 c^2 w^4)}{ 2 \rho \omega^4 }.
        \label{P_critical}
    \end{equation}
    The above expression indicates that the critical power varies with the propagation direction $z$. This variation directly follows the diffraction profile because of its linear dependence on $\mu(z)$, which appears on the right-hand side of the equation emphasizes this relationship. 

    Further by differentiating the Eq.~(\ref{beamwidth}) with respect to $z$, and normalizing the result by using $y(z)= 
    \omega / \omega_0$ [where $\omega_0 = \omega(0)$] we get, force

    \begin{equation}
        F(y)=\frac{1}{\mu} \frac{d^2  y}{d z^2}= \frac{\mu}{\omega_0^4 y^3} + \frac{2 c y}{\mu} \frac{d \mu}{d z}  - \frac{2 \rho P_0 y}{\sqrt\pi \left( \sigma^2 + 2 \omega_0^2 y^2\right)^{3/2}}.
        \label{force}
     \end{equation}
    The above equation can be interpreted as the equation of motion for a one-dimensional particle whose effective mass varies with position as $\frac{1}{\mu(z)}$. Therefore, the force acting on a Gaussian beam in an inhomogeneous SNNM can be described by the force on a particle with variable mass. With reference to the work by Qi Guo et al. in \cite{guo2004large}, the first and second terms in Eq. (\ref{force}) arise due to diffraction, which accounts for the expansion of the beam. The third term in Eq. (\ref{force}) arises due to the refractive force generated by the nonlinearity induced refraction. As a result of this force, the beam undergoes a compression effect. However, when these two forces are in balance, the formation of a stable soliton occurs, as in the case of constant diffraction. We note here that if $\mu$ is constant then the corresponding beam dynamics in homogeneous nonlocal medium is governed by the force $F(y)=\frac{\mu}{\omega_0^4y^3}-\frac{2 \rho P_0 y}{\sqrt\pi \left( \sigma^2 + 2 \omega_0^2 y^2\right)^{3/2}}$ with constant effective mass $1/\mu$ \cite{guo2004large}.  

    The corresponding form of potential $V(y)$ is identified as 
    \begin{multline}
          V\left( y \right)= \frac{\mu \left( 1- y^2 \right)}{2 \omega_0^4 y^2} - \frac{P_0 \rho}{\sqrt{\pi} \omega_0^2}\left[ \frac{1}{\sqrt{ \sigma^2 + 2 y^2 \omega_0^2}} -  \frac{1}{\sqrt{ \sigma^2 + 2 \omega_0^2}}\right] \\ - \frac{c \left(1 -y^2\right)}{\mu} \frac{d \mu}{d z}.
        \label{petential}
     \end{multline}

         \section{Beam propagation dynamics}

   To investigate the propagation dynamics of the Gaussian beam in inhomogeneous media and to tailor the characteristics of accessible solitons, we have solved the coupled ODEs (Eqs. (\ref{ampitude})-(\ref{phase})) by using the classical RK4 method. The initial parameter values were chosen as $c = 0$, $\omega = 1$, $\rho = 1/6$, and $\sigma = 10$. The other parameter values, such as the constant diffraction value $\mu_0$ and the critical power $P_c$ are fixed at $0.3$ and $1643.3$ for the cases of constant, linear and exponentially decreasing diffraction. For the other three cases, namely step-like, barrier-type, and periodic diffraction modulations, these values are fixed at $1.25$ and $6847.1$. With these initial conditions, the coupled ODEs were solved for three input powers, namely $P_0 < P_c$, $P_0 = P_c$, and $P_0 > P_c$. We remark here that the critical power can be chosen depending on whether a low- or high-beam operation is required.  

\begin{figure*}[hbt!]
        \centering
             \includegraphics[scale=0.44]{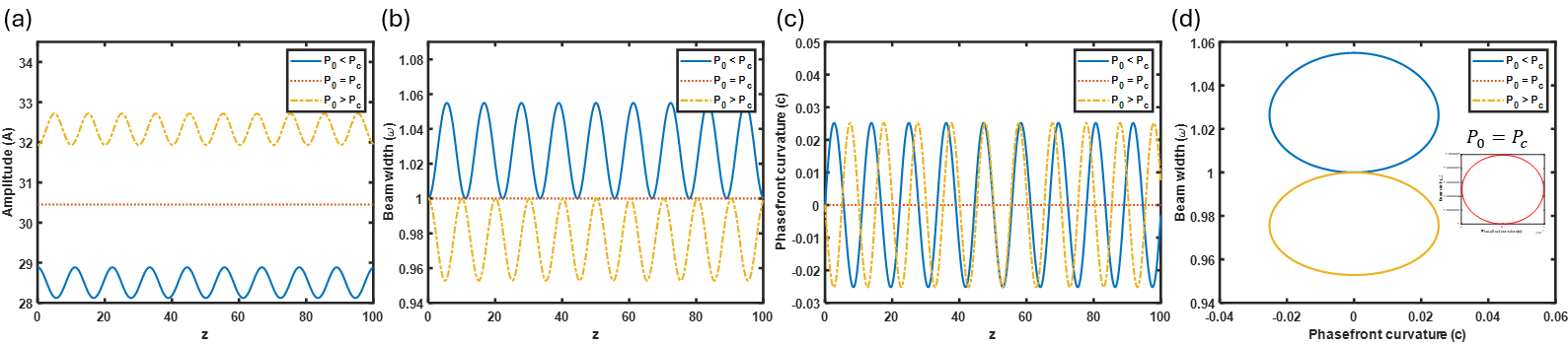}
               \caption{Beam characteristics for constant diffraction $\mu(z)=\mu_0=0.3$:  (a) Amplitude, (b) Beam width, (c) Phase-front curvature, (d) Phase space trajectory, shown  for three different power conditions: $P_0 < P_c$ (blue line), $P_0 = P_c$ (red dotted line) and $P_0 > P_c$ (yellow dashed-dotted line).}
         \label{Constant}
   \end{figure*}

   In order to obtain a clear understanding of the dynamics of nonlocal solitons in the presence of variable diffraction, it is essential to analyze the nature of accessible solitons in the absence of inhomogeneity. That is, for constant diffraction $\mu(z)=\mu_0=0.3$. Under this choice, the Gaussian beam parameters, such as amplitude, width, curvature, and phase, do not vary along the longitudinal direction and are remain constant. The intensity of the Gaussian-shaped solution ($\psi=\frac{\sqrt{P_0}}{(\sqrt{\pi}\omega)^{1/2}}e^{-\frac{x^2}{2\omega^2}}e^{i(cx^2+\theta)}$) is expressed as $|\psi|^2=\frac{P_0}{\sqrt{\pi}\omega}e^{-\frac{x^2}{\omega^2}}=A_0^2e^{-\frac{x^2}{\omega^2}}$. The corresponding beam characteristics are presented in Fig. \ref{Constant}. Figure \ref{Constant}(b) shows that for $P_c>P_0$, the beam width initially expands from 1 to a maximum value of 1.06, and then decreases to a minimum of 1. This oscillatory behavior continues along the propagation direction, leading to the formation of breathers. It occurs because the diffractive force dominates over the nonlinearity-induced refractive force. We have confirmed this breathing dynamics  in Fig. \ref{ConstantSSFT}(a) by performing numerical simulations of the corresponding beam evolution using the SSFT method.  On the other hand, when $P_c<P_0$, the refractive force dominates over the diffractive force, causing the beam to contract to a minimum value of $0.95$ and then expand back. As a result, the beam width oscillates between $1$ and $0.95$ along the $z$-direction, leading to the formation of breathers. The corresponding beam evolution is displayed in Fig. \ref{ConstantSSFT}(c). However, a stable AS forms when $P_c=P_0$, as indicated by the dotted red line in Fig. \ref{Constant}. In this situation, the amplitude, beam width, and the corresponding phase-front curvature do not oscillate, as shown in Figs. \ref{Constant}(a), \ref{Constant}(b), and \ref{Constant}(c), respectively. In Fig. \ref{Constant}(d), we show the beam width-curvature ($\omega-c$) phase diagram, from which the appearance of closed loops for all three power conditions confirms the stability of the nonlocal soliton.Numerically simulated stable beam evolution for this case is displayed in Fig. \ref{ConstantSSFT}(b), which ensures that the AS propagates without distortion at $P_c=P_0$.
 
   \subsection{\textbf{Beam propagation in nonlocal media with linearly and exponentially decreasing diffraction}}
\begin{figure*}[hbt!]
        \centering
             \includegraphics[scale=0.44]{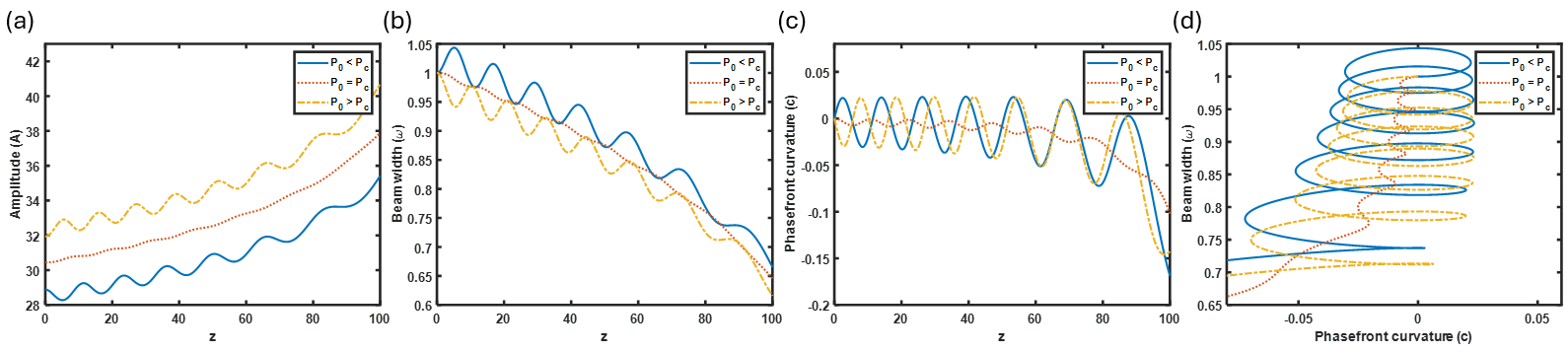}
               \caption{Beam characteristics for linear diffraction $\mu(z)=\mu_0 [\frac{1-\beta}{\beta L}z +  1]$:  (a) Amplitude, (b) Beam width, (c) Phase-front curvature, (d) Phase space trajectory, shown  for three different power conditions: $P_0 < P_c$ (blue), $P_0 = P_c$ (red) and $P_0 > P_c$ (yellow).}
         \label{Linear}
   \end{figure*}

   To study the effect of longitudinally-varied diffraction on the evolution of a Gaussian beam, we consider the Gaussian ansatz as given in (Eq.~\ref{ansatz}), in which the parameters now depend on $z$. We begin by explaining the evolution of a Gaussian beam under the influence of a linearly decreasing diffraction profile (Eq. (\ref{linear-diffraction})) with the initial parameter  values $\beta=6$ and $L=100$. As discussed above, it is a well-known fact that in the constant diffraction case, breather formation occurs only when $P_c>P_0$ or $P_c<P_0$, although the underlying mechanisms differ.  For $P_c=P_0$, a stable soliton state is formed. However, in the present linearly varying diffraction case, the Gaussian beam evolves into a breather-like state under all three power conditions, as shown in Fig. \ref{Linear}. No soliton formation occurs with zero oscillations for $P_c=P_0$, unlike in the constant diffraction case. An interesting observation is that the mechanism responsible for the formation of breathers in the linearly decreasing diffraction case for $P_c>P_0$ and $P_c<P_0$ is identical to that in the constant diffraction case under the same power conditions. However, the mechanism that leads to the breather-like state observed at $P_c=P_0$ is distinctly different. This means that the diffractive and refractive forces in Eq. (\ref{force}) are no longer balanced. As a result, the beam's width expands and shrinks alternately because the diffractive force becomes stronger than the refractive one. This behavior is shown in Fig. \ref{Linear}(b) by red dotted line. It is worth noting that the amplitude of beam's width oscillation is relatively small compared to that observed under the other power conditions. In addition to these observations, as shown in Fig. \ref{Linear}(b), the beam width decreases progressively along the $z$-direction. This is the result of the linear decrease of diffraction along the propagation direction, which leads to an increase in amplitude and a corresponding decrease in beam width, which can also be confirmed from the beam evolution Fig. \ref{LinearSSFT}(b). To ensure this, one can calculate the full-width at half maximum from the intensity expression of the Gaussian solution (\ref{ansatz}). In addition, the beam width exhibits higher periodicity at the initial stages, which gradually reduces with increasing $z$. A similar behavior is observed in the amplitude and the chirp.

    Further, Fig. \ref{Linear}(a) shows that the beam amplitude exhibits continuous growth with oscillations for $P_0 > P_c$ and $P_0<P_c$, while at $P_0 = P_c$, the amplitude increases monotonically with only slight oscillatory modulation.
    Similarly, the phase-front curvature in Fig.~\ref{Linear}(c) exhibits pronounced oscillations with large amplitudes when $P_0 \neq P_c$, in contrast to the comparatively weaker but visible oscillatory behaviour at $P_0 = P_c$. In both cases, however, the overall trend is towards broadening as the beam propagates.  The corresponding $\omega-c$ phase diagram is shown in Fig. \ref{Linear}(d) to further emphasize the influence of the linearly decreasing diffraction profile. At all power conditions, the trajectories do not form a completely closed loop. Instead, they tend to approach closure but begin tracing a new loop before completing a full cycle. This behavior arises from the modulation induced by the linearly decreasing diffraction. Overall, the beam width exhibits periodic oscillations, leading to the formation of a diffraction-managed breather soliton in SNNM, whose profile is further modulated according to the linear diffraction variation. The numerically simulated diffraction-managed breather solitons for the subcritical and supercritical cases are displayed in Fig. \ref{LinearSSFT}(a) and Fig. \ref{LinearSSFT}(c), respectively.

     \begin{figure*}[hbt!]
        \centering
            \includegraphics[scale=0.6]{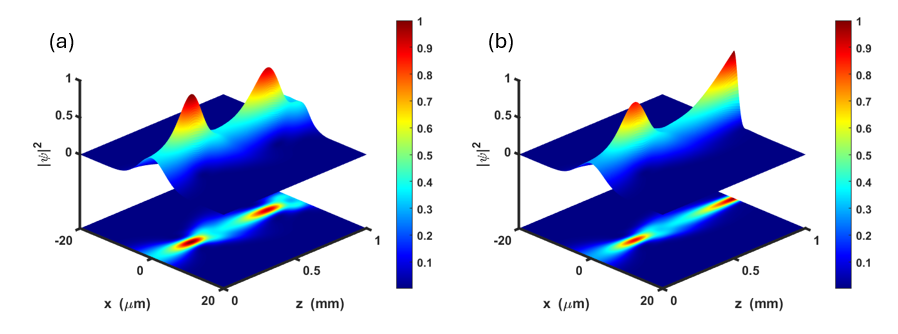}
            \caption{Numerical simulation of the evolution of a Gaussian beam in the SNNM at \( P_0 >P_c \) with a  (a) Constant diffraction, (b) For linearly decreasing diffraction}
        \label{SSFTrealParameter}
  \end{figure*}

    We now analyze the case of exponentially decreasing diffraction [Eq.~(\ref{exponential-diffraction})]. The effect of this modulation is qualitatively similar to that observed in the case of linearly decreasing diffraction (therefore, for the sake of brevity, we do not provide the figures of beam characteristics and the corresponding beam evolution figures). In particular, the beam amplitude increases persistently along the propagation distance, accompanied by oscillations for \(P_0 \neq P_c\), while at the critical power \(P_0 = P_c\) the amplitude increases smoothly with only weak oscillatory modulation. The corresponding beam width evolution exhibits continuous oscillations analogous to those observed in the linear diffraction case, together with a gradual contraction of the beam but the period of the oscillation is higher than that of the linear case. An additional notable feature is that the oscillation period is initially shorter and gradually increases as the beam propagates along the \(z\)-direction. Similar behaviour is observed in the evolution of the beam 
    amplitude and the phase-front curvature.
    The phase-front curvature displays oscillatory dynamics with an overall broadening trend for \(P_0 \neq P_c\), whereas at the critical power \(P_0 = P_c\) it broadens with significantly reduced oscillation amplitude, even though the overall dynamic is similar to the linear case, the broadening is more as compared to the linear case. Most importantly, the phase-space trajectories exhibit spiral-like motion for all considered input powers, confirming the formation of diffraction-managed breather solitons. The characteristics of these breathers are influenced by the exponential diffraction modulation: for \(P_0 \neq P_c\), the spirals are wider, indicating stronger oscillatory behaviour, while at \(P_0 = P_c\) the trajectory gradually contracts, demonstrating that exponential inhomogeneity supports breather formation even in the critical case. 

    Overall, the strongly nonlocal medium supports breather-like soliton evolution under both linearly and exponentially decreasing diffraction inhomogeneities. The mechanism of breather soliton formation and the resulting profile modulation follow more or less the same pattern in both cases.

    To demonstrate that the obtained breather dynamics in the cases of constant diffraction and linearly varying diffraction, under supercritical case, can also be observed experimentally, we adopt the parameter values considered in Ref. \cite{conti2004observation,peccianti2000electrically}.  In that work, pulsating solitons were experimentally observed in a NLC planar cell by considering an operating wavelength of $\lambda = 1064\,\mathrm{nm}$, a Gaussian beam waist of $\omega_{0} \approx 5\,\mu\mathrm{m}$, and a propagation distance of $z \approx 1\,\mathrm{mm}$ as our initial conditions in our numerical analysis. The input optical power is taken to be $P \approx 6.89\,\mathrm{mW}$, corresponding to the milliwatt-level powers commonly required to induce reorientational nonlinearity in NLC planar medium. Further, the value of the nonlocal response length considered in our simulations is $\sigma = 10$, and the assumed modulation strength is $\mu_0 = 0.0529$. The numerical results are presented in Figs. \ref{SSFTrealParameter}(a) (for constant diffraction) and \ref{SSFTrealParameter}(b) (for linearly decreasing diffraction), which clearly agree with the experimentally observed pulsating soliton shown in Fig. 1(b) of Ref. \cite{conti2004observation}. These numerical experiments confirm that the theoretically predicted breathing solitons are, in principle, experimentally realizable.
    
    \subsection{\textbf{Beam propagation in nonlocal media with step-like, barrier-type and periodic diffraction modulations}}
   
    We now consider three other physically important forms of longitudinally varying diffraction modulations that are correlated with existing experiments in local Kerr media. In particular, we consider step-like, barrier-type, and periodic variations along the propagation direction, described by Eqs. (\ref{tanh}), (\ref{sech}), and (\ref{periodic}), respectively. 

\begin{figure*}[hbt!]
        \centering
        \includegraphics[scale=0.44]{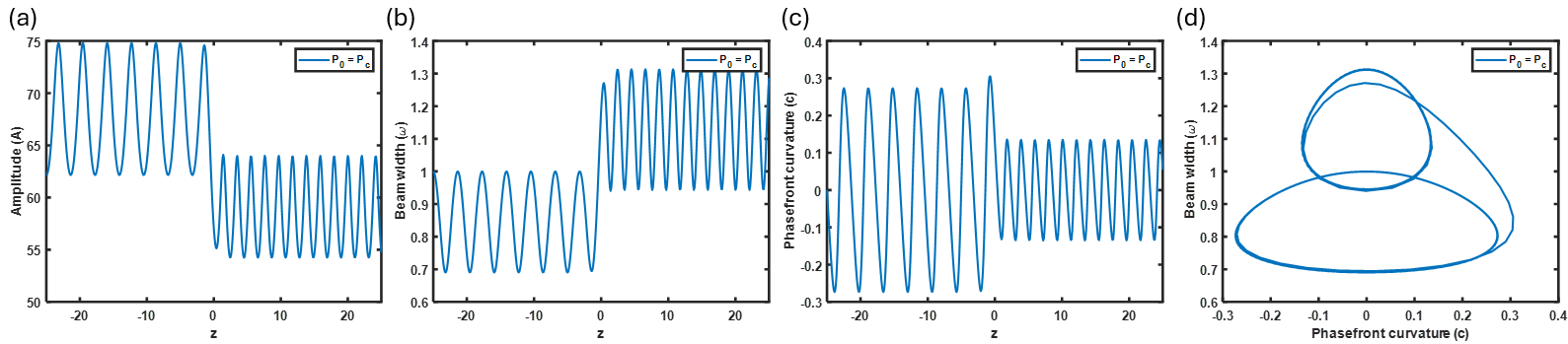}
            \caption{Step-like modulation of diffraction profile $\mu(z) = \mu_0 +\mu_1  \tanh(\mu_2 z + \mu_3)$ on the beam dynamics is shown for the power $P_0 = P_c$. (a) Amplitude, (b) Beam width, (c) Phasefront curvature, (d) Phase space trajectory. The constants are $\mu_0=1.25$, $\mu_1=0.65$, $\mu_2=1.25$ and $\mu_3=0.05$. }
        \label{Tanh}
    \end{figure*}

    Under the influence of a step-like modulation of diffraction, modelled by the hyperbolic tangent profile (see Fig. \ref{diffraction profile}(b)), we observe qualitatively distinct beam dynamics compared to the linearly and exponentially decreasing diffraction cases. A typical behavior of this diffraction modulation is that the medium exhibits a specific diffraction property up to a certain propagation distance and then undergoes a sudden change at $z=0$. Overall, the diffraction profile along the $z$-direction exhibits a kink-like nature. Due to this diffraction modulation, the beam undergoes a step-like variation in the soliton evolution, as shown in Fig.~\ref{Tanh} for $P_c=P_0$ (for brevity, we have omitted the details of $P_c\neq P_0$ because of a similar effect of modulation). In particular, the beam width exhibits a smooth step-like modulation combined with periodic oscillations as depicted in Fig. \ref{Tanh}(b). We observe that for $z<0$, the beam undergoes continuous contraction followed by expansion, resulting in a breather soliton with a period $T_1=2\pi/\Omega_1$. During the transition point at $z=0$, the beam width suddenly expands to its maximum value and then contracts in the opposite direction, reaching a minimum value of $1$. The oscillations persist beyond this transition point with a new period $T_2=2\pi/\Omega_2$. That is, for $z>0$, the beam width experiences broadening accompanied by an increase in periodicity. 
    Consequently, the breather soliton begins to oscillate with a new frequency $\Omega_2$ for $z>0$. Then, as far as the amplitude of this breathing soliton is concerned, it undergoes a sudden suppression and exhibits an oscillatory behaviour. Notably, the oscillations are stronger for $z>0$ compared to those observed for $z<0$. 
    The phase-front curvature corresponding to this case is demonstrated in Fig. \ref{Tanh}(c), in which the curvature exhibits an oscillatory behaviour throughout the propagation direction, but at $z=0$ a noticeable transition occurs. Importantly, the presence of closed loops associated with two distinct stable equilibrium points in the phase-space trajectory, as illustrated in Fig. \ref{Tanh}(d), confirms that the system supports bounded and stable breather-type soliton dynamics in two distinct regions of the propagation distance.  To further verify the beam dynamics under step-like diffraction modulation, we numerically simulated the corresponding beam evolution, as shown in Fig.~\ref{SSFTprriodicTanhSech}(a). This figure further corroborates the features described above for the longitudinally varying step-like diffraction modulation.
    
\begin{figure*}[hbt!]
        \centering
        \includegraphics[scale=0.44]{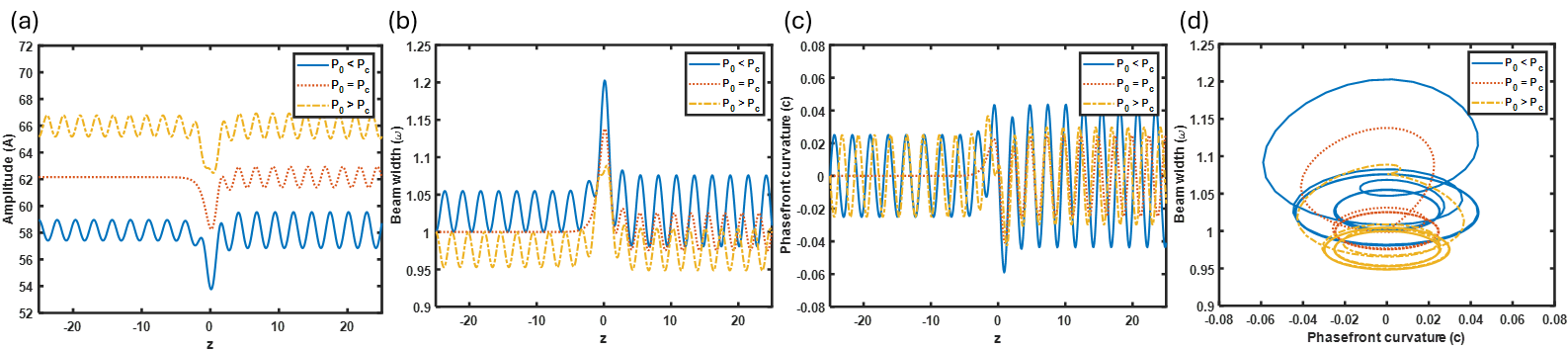}
           \caption{ Influence of the barrier-type diffraction profile $\mu(z) = \mu_0 + \mu_1 ,\sech(\mu_2 z + \mu_3)$ on the beam dynamics, illustrated through (a) amplitude, (b) beam width, (c) phase-front curvature, and (d) phase-space trajectory for three power $P_0<P_c$ (blue line), $P_0 = P_c$ (red line), $P_0>P_c$ (yellow line). The parameter values are the same as in the step-like diffraction modulation.}
        \label{Sech}
    \end{figure*}

   Now, we consider a strongly nonlocal medium with a localized barrier-type diffraction modulation described by Eq. (\ref{sech}). This consideration addresses an interesting question in beam dynamics: how does the Gaussian beam evolve in the presence of a localized inhomogeneity? In contrast to the tanh case, this $\sech$ diffraction modulation imparts intriguing and distinctive effects on the beam dynamics. As shown in Fig. \ref{diffraction profile}(b), the peak of the $\sech$-function can be positioned anywhere on the $z$-axis by adjusting the constants $\mu_j$'s, $j = 0, 1, 2, 3$, allowing precise control over the location of the localized inhomogeneity in the waveguide. Due to this modulation, as we have demonstrated in Fig. \ref{Sech}(b), the beam width exhibits oscillations in the negative $z$-direction for $P_0 \neq P_c$. At $z = 0$, the beam width undergoes a sudden expansion followed by contraction. For $z > 0$ and under the same power condition ($P_0 \neq P_c$), it continues to oscillate, but with an increased amplitude. Therefore, under the power condition $P_0 \neq P_c$, the localized inhomogeneous diffraction induces large-amplitude oscillations. Another notable observation is that the beam is not reflected but transmitted. However, when $P_0 = P_c$, the beam neither expands nor contracts and maintains a stable profile until it reaches $z = 0$. At $z = 0$, the beam width undergoes a sudden expansion, followed by oscillations. This behavior indicates that the localized inhomogeneity induces oscillations as the beam passes through it. 

   In Fig. \ref{Sech}(a), we illustrate the effect of the barrier-type diffraction on the amplitude of a Gaussian beam for both $P_0 \neq P_c$ and $P_0 = P_c$. For $P_0 = P_c$, the amplitude remains constant for $z < 0$, while for $P_0 \neq P_c$, oscillations are present in the negative $z$-direction. In both cases, the soliton amplitude is suppressed at $z = 0$ due to the sudden expansion of the beam width. Beyond this critical point, the amplitude of the oscillations increases slightly. A similar scenario is observed in Fig. \ref{Sech}(c) for the phase-front curvature as well. The phase diagram shown in Fig. \ref{Sech}(d) indicates that the localized inhomogeneity induces a diffraction-managed breathing soliton for $P_0 \neq P_c$. In contrast, for $P_0 = P_c$, the beam propagates as a stable, smooth soliton before encountering the inhomogeneity. After passing through the localized inhomogeneity, it transforms into a stable breather soliton, as depicted in Fig. \ref{SSFTprriodicTanhSech}(b).

\begin{figure*}[hbt!]
        \centering
            \includegraphics[scale=0.44]{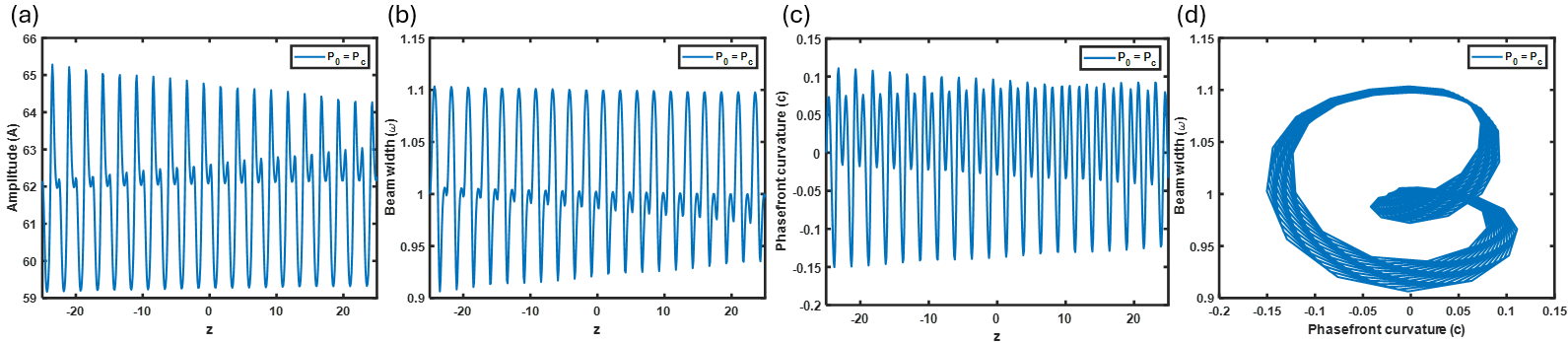}
           \caption{The role of the periodic diffraction profile $\mu(z) = \mu_0 + \mu_1 \sin(\mu_2 z + \mu_3)$ on the beam dynamics, illustrated through (a) amplitude, (b) beam width, (c) phase-front curvature, and (d) phase-space trajectory for the power $P_0 = P_c$. The parameter values are the same as in the previous step-like case, except $\mu_2 = 5$.}
        \label{Periodic}
    \end{figure*}

\begin{figure*}[hbt!]
       \centering
          \includegraphics[scale=0.55]{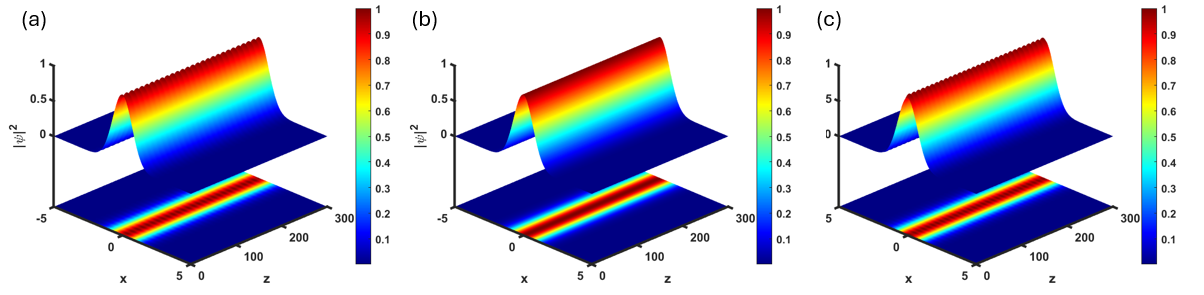}
            \caption{Evolution of a Gaussian beam in the SNNM with a constant diffraction profile for (a) \( P_0 < P_c \), (b) \( P_0 = P_c \), and  (c) \( P_0 > P_c \). }
       \label{ConstantSSFT}
    \end{figure*}

\begin{figure*}[hbt!]
       \centering
          \includegraphics[scale=0.55]{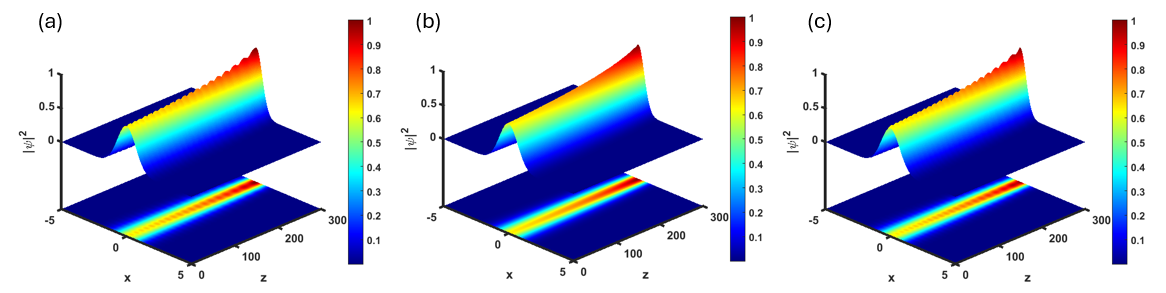}
          \caption{Evolution of a Gaussian beam in the SNNM with a linearly decreasing diffraction profile for  (a) \( P_0 < P_c \), (b) \( P_0 = P_c \), and  (c) \( P_0 > P_c \)}
       \label{LinearSSFT}
    \end{figure*}

\begin{figure*}[hbt!]
       \centering
          \includegraphics[scale=0.55]{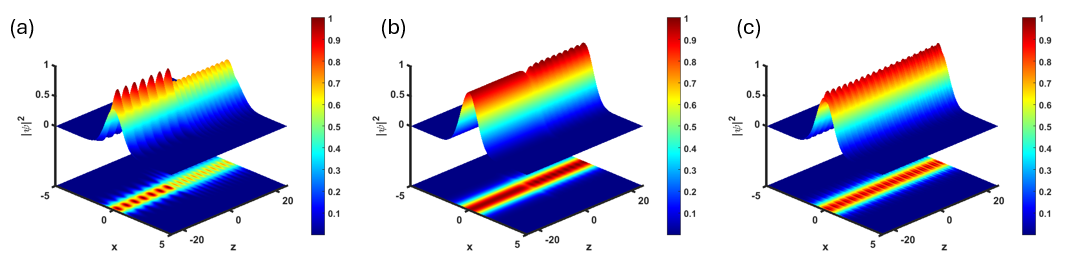}
           \caption{The evolutions of a Gaussian beam in SNNM obtained using the SSFT method with different diffraction profiles are displayed: (a) step-like: $\mu(z)= \mu_0 +\mu_1  \tanh(\mu_2 z + \mu_3)$, (b) barrier-type: $ \mu(z) = \mu_0 +\mu_1  \sech(\mu_2 z + \mu_3)$, and (c) periodic: $ \mu(z) = \mu_0 + \mu_1  \sin(\mu_2 z + \mu_3)$.   }
       \label{SSFTprriodicTanhSech}
    \end{figure*}

    Finally, we investigate the beam dynamics under a periodic modulation of the diffraction denoted by Eq. (\ref{periodic}). The resulting beam characteristics for $P_c=P_0$ are displayed in Fig. \ref{Periodic} (here we have omitted the results for $P_c\neq P_0$ for brevity). Unlike the previously analyzed cases, the periodic diffraction induces a breathing behavior in the beam parameters along the propagation axis. This results in the formation of a breathing nonlocal soliton, as shown in Fig. \ref{SSFTprriodicTanhSech}(c). We have illustrated the corresponding  amplitude, beam width, and phase-front curvature in Figs.~\ref{Periodic}(a)-(c), respectively.  These figures reveal that all three beam parameters exhibit oscillatory behavior with clear periodicity, effectively reflecting the periodic nature of the underlying diffraction modulation. Most importantly, the phase-space trajectory [Fig.~\ref{Periodic}(d)] forms closed curves, confirming the bounded and stable nature of the soliton dynamics. This behavior signifies the existence of stable breather-type solitons supported by the periodic diffraction modulation.
    
 The summary of the evolution and characteristic behaviour of the beam parameters analyzed in this study is presented in Table~\ref{Table}.
 
\begin{table*}[h!]
\centering
\scriptsize
\begin{tabular}{|p{1.2cm}|p{1.9cm}|p{1.9cm}|p{1.9cm}|p{1.9cm}|p{1.9cm}|p{1.9cm}|p{1cm}|p{1cm}|}
\hline
\textbf{Diffraction Profile} & \multicolumn{2}{c|}{\textbf{Amplitude}} & \multicolumn{2}{c|}{\textbf{Beam width}} & \multicolumn{2}{c|}{\textbf{Phasefront curvature}} & \textbf{Phase diiagram} & \textbf{Beam evolution} \\ 
\cline{2-7}
 & $P_0 = P_c$ & $P_0 \neq P_c$ & $P_0 = P_c$ & $P_0 \neq P_c$  & $P_0 = P_c$ & $P_0 \neq P_c$  &  &\\ 
\hline
\textbf{Constant} & Stable & Oscillating & Stable & Oscillating & Stable & Oscillating & Closed loop & Soliton $(P_0 = P_c)$  Breather $(P_0 \neq P_c)$\ \\
\hline
\textbf{Linearly Decreasing} & Monotonically increases  with very small oscillation & Monotonically increases with pronounced oscillation & Monotonically increases with small oscillation & Monotonically increases with pronounced oscillation & Broadening with oscillation & Broadens more with oscillation & Spiral & Breather  \\ 
\hline
\textbf{Exponentially Decreasing} & Monotonically increases with very small oscillation & Monotonically increases with very pronounced oscillation & Monotonically increases with very small oscillation & Monotonically increases with very pronounced oscillation & Broadening with oscillation & Broadens more with oscillation & Spiral & Breather\\ 
\hline
\textbf{Step like} & Oscillating – Sudden jump at $z = 0$. Increased oscillation for $z > 0$ & Oscillating – Sudden jump at $z = 0$. Increased oscillation for $z > 0$ & Oscillating – Sudden increase at $z = 0$. Increased oscillation for $z > 0$ & Oscillating – Sudden increase at $z = 0$. Increased oscillation for $z > 0$ & Curvature is broader in the region ($z < 0$) and it narrows down for $z > 0$ & Curvature is broader in the region ($z < 0$) and it narrows down for $z > 0$  & Closed loop & Breather\\ 
\hline
\textbf{Barrier type} & Stable propagation for $z < 0$. Oscillating $z > 0$. There is a dip at $z = 0$ & Oscillating for both $z < 0$ and $z > 0$ region. There is a dip at $z = 0$ & Stable propagation for $z < 0$. Oscillating $z > 0$. There is a spike at $z = 0$ & Oscillating for both $z < 0$ and $z > 0$ region. There is a spike at $z = 0$ & Stable for $z < 0$, oscillating for $z > 0$ & Oscillating for both $z < 0$ (less amplitude) and $z > 0$(more amplitude) & Closed loop  & Breather \\ 
\hline
\textbf{Periodic} & Periodically oscillating with slight decreases in amplitude along $z$ & Periodically oscillating & Periodically oscillating with slight decreases along $z$ & Periodically oscillating & Periodically oscillating  & Periodically oscillating & Closed loop & Breather\\ 
\hline

\end{tabular}
\caption{Summary of Gaussian beam dynamics for different diffraction profiles.}
\label{Table}
\end{table*}

    In the constant-diffraction regime, stationary stable solitons form at the critical power in most nonlinear media (local cubic Kerr, cubic-quintic, weakly nonlocal, nonlocal logarithmic, and strongly nonlocal media), while different dynamical behaviors (such as diffraction, collapse, and the formation of breathing and quasi-breathing solitons) arise in the subcritical and supercritical regimes depending on the type of nonlinearity \cite{freitas1998propagation, keshavarz2016gaussian, jovanoski1994propagation, karlsson1992optical, wang2018nonlocal, tsoy2010solitons, mishra2022generation}. However, when variable diffraction is introduced, particularly in strongly nonlocal nonlinear media, stable solitons transform into breathing or oscillatory structures. Thus, the proposed diffraction-management scheme enables effective control of beam dynamics in all power regimes by appropriately tailoring the diffraction profile and input power.

    Following the studies on the propagation characteristics of the nonlocal soliton, in the subsequent subsection, we analyze the influence of the strength of nonlocality on the beam dynamics as it decreases toward the weakly and generally nonlocal regimes. 

\subsection{Beam dynamics with the variation of the strength of the nonlocality}

\begin{figure*}[hbt!]
        \centering
            \includegraphics[scale=0.55]{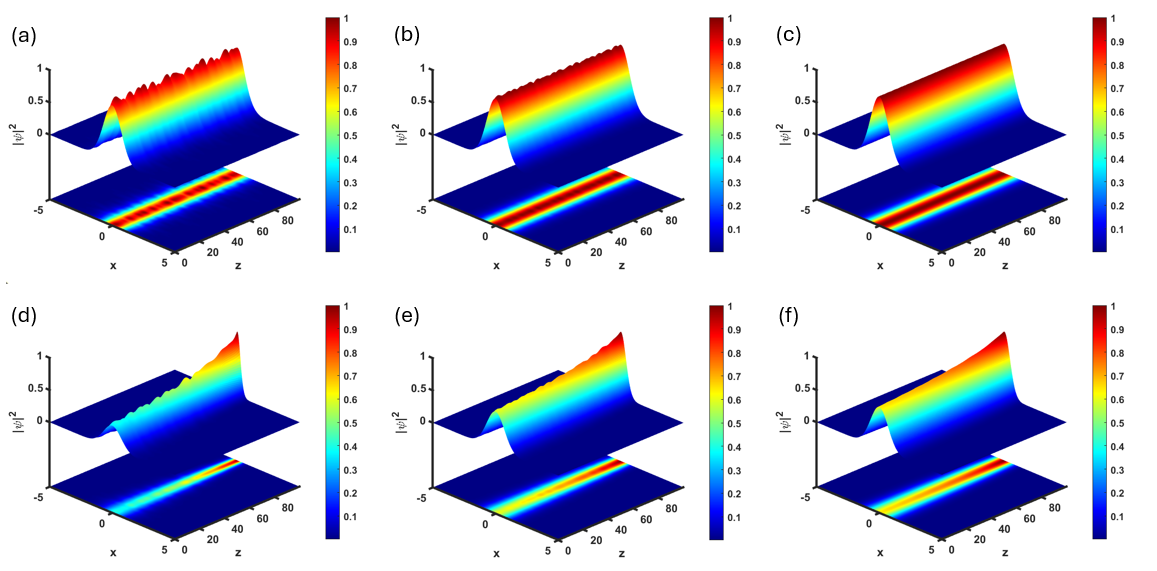}
            \caption{The variation of evolutions of a Gaussian beam with the strength of the nonlocality. First, second and third columns are for weakly nonlocal ($\sigma=0.5$), generally nonlocal($\sigma = 2$) and for strongly nonlocal($\sigma = 10$) cases, respectively. the first row is for the constant diffraction, the second row is for linearly decreasing diffraction.}
        \label{ConLinSigma}
    \end{figure*}

\begin{figure*}[hbt!]
        \centering
            \includegraphics[scale=0.55]{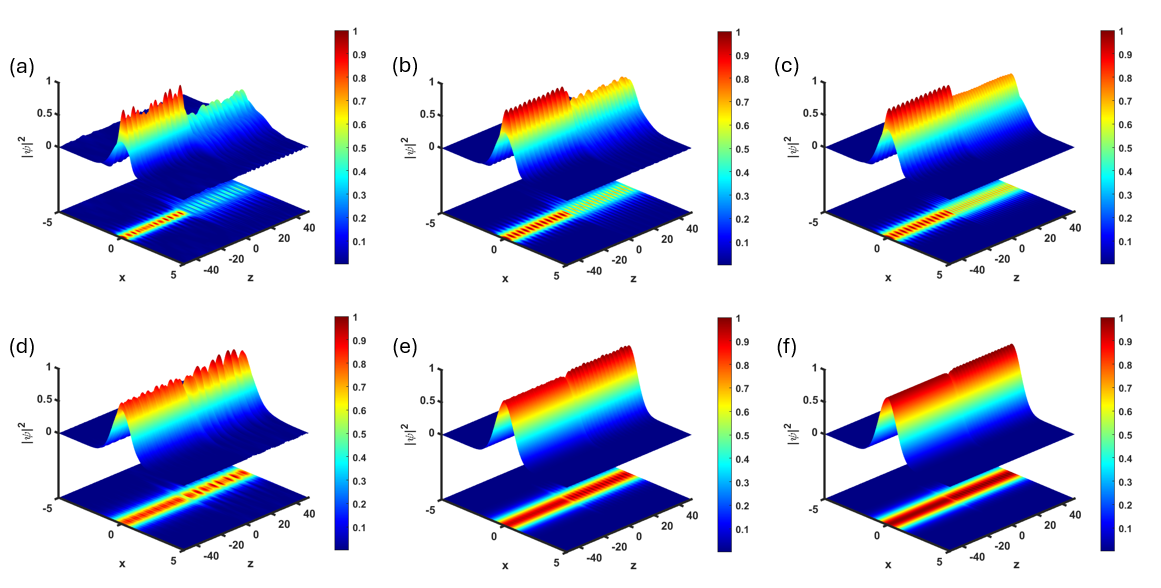}
            \caption{The variation of evolutions of a Gaussian beam with the strength of the nonlocality. First, second and third columns are for weakly nonlocal ($\sigma=0.5$), generally nonlocal($\sigma = 2$) and for strongly nonlocal($\sigma = 10$) cases, respectively. The first row is for the step-like diffraction, the second row is for Barrier-type diffraction.}
        \label{TanhSechSigma}
    \end{figure*}

In order to examine the role of the nonlocality strength parameter $\sigma = L_c$ on Gaussian beam dynamics, we choose the initial beam width ($\omega_0$) relative to a given value of $\sigma$ such that the conditions corresponding to the different nonlocal regimes, as described in the Introduction, are satisfied. The results of our numerical analysis are presented in Figs. ~\ref{ConLinSigma} and \ref{TanhSechSigma} . In particular, we present numerical results for the constant diffraction case, as well as for linear, step-like, and barrier-type modulated diffraction profiles, under the condition $P_0=P_c$. The figures clearly demonstrate the existence of stable solitons in all diffraction cases in the strongly nonlocal limit ($\sigma=10$). When the strength of nonlocality is reduced from $\sigma=10$ to $\sigma=2$ (comparable to the beam-width value $\omega_0=1$), fluctuations start to appear in the structure of the nonlocal soliton, as shown in Figs. \ref{ConLinSigma}(b) \& \ref{ConLinSigma}(e) and \ref{TanhSechSigma}(b) \& \ref{TanhSechSigma}(e). We refer to this solitonic state as a quasi-stable soliton, as the terminology is commonly used in the nonlocal literature. Then, when the nonlocality strength is further reduced to $\sigma=0.5$, the disturbances become more visible, as depicted in Figs. \ref{ConLinSigma}(a) \& \ref{ConLinSigma}(d) and \ref{TanhSechSigma}(a) \& \ref{TanhSechSigma}(d). Consequently, the quasi-soliton structure undergoes instability. This implies that the medium’s ability to counterbalance diffraction gradually weakens, leading to instability. From this numerical analysis, we conclude that strongly nonlocal media support stable diffraction-managed breathing solitons, while weakly and generally nonlocal media give rise to comparatively less stable solitonic structures.

Then, in the next section,  we investigate the stability of the soliton by performing modulation instability studies through linear stability analysis.

  \section{MODULATIONAL INSTABILITY}
    Modulational instability is the spontaneous amplification of perturbations in the amplitude or phase of a continuous wave as it propagates through a nonlinear medium. It typically arises when the delicate interplay between diffraction (or dispersion) and nonlinearity results in the self-induced formation of sidebands in the wave spectrum \cite{agrawal2000nonlinear}. It has been proven that spatial nonlocality can significantly impact MI features like the gain and bandwidth \cite{krolikowski2001modulational, krolikowski2004modulational,tabi2022modulational}. For instance, it has been reported that, in the case of constant diffraction, the MI cannot be completely eliminated for a focusing nonlinearity ($\rho>0$ in Eq. (\ref{NNLSE})), although strong nonlocality tends to suppress it \cite{krolikowski2001modulational}. In the case of defocusing nonlinearity ($\rho < 0$), the MI depends on the profile of the response function. For a smooth Gaussian response, the plane wave remains stable, and MI does not occur. However, MI may arise when the response function has a rectangular profile \cite{krolikowski2001modulational}. Ultimately, the investigation of how variable diffraction affects the MI dynamics is of particular interest. In this section, we address this issue by analyzing the five distinct diffraction profiles mentioned earlier.

    Here  we are going to study MI for the self focusing $(\rho >0)$ only. It can be extended to the self defocusing $(\rho <0)$ case too. For this purpose, let us consider the plane wave solution of Eq. (\ref{NNLSE}) as follows
    \begin{equation}
      \psi(z,x) = \sqrt p_0 e^{i\varphi},~\varphi=k_0 x - \omega_0' z, \label{planewave}
    \end{equation}
    where $\sqrt{p_0}$ corresponds to the amplitude of the plane wave and  $k_0$ and $\omega_0'$ denote its wave number and frequency, respectively. These are linked through the nonlinear dispersion relation 
    \(\omega_0 = \frac{\mu(z) k_0^2}{2} - \rho p_0\). Note that the variable diffraction explicitly appears in the dispersion relation. 

    To study the stability of the plane wave (\ref{planewave}) we introduce a small complex perturbation $a(z,x)=u(z,x)+iv(z,x)$ in Eq. (\ref{planewave}) as follows
    \begin{equation}
       \psi(z,x) = [\sqrt{p_0}+u(x,z)+iv(x,z) ]\exp(i(k_0 x - \omega_0' z)),
       \label{perturbed_field}
    \end{equation}
    where $u(x,z)$ and $v(x,z)$ are the real and imaginary parts of the complex perturbation $a(x,z)$, respectively. Substituting \eqref{perturbed_field} into the NNLSE (\ref{NNLSE}), and followed by linearization around the unperturbed solution (\ref{planewave}), we obtain the following pair of coupled equations for the perturbation after separating it into real and imaginary parts:

   \begin{subequations}
       \begin{align}
        \frac{\partial u}{\partial \tau} &= -\frac{\mu(\tau)}{2} \frac{\partial^2 v}{\partial \xi^2} \label{coupled1} \\
        \frac{\partial v}{\partial \tau} &= \frac{\mu(\tau)}{2} \frac{\partial^2 u}{\partial \xi^2} + 2\rho\, p_0 \int_{-\infty}^{\infty} u(\xi', \tau ) R(\xi - \xi') \, d \xi'. \label{coupled2}
    \end{align}
   \end{subequations}
    Here, new coordinates $\tau$ and $\xi$ are defined by $\tau = z$ and $\xi = x - \mu k_0 z $.  By following Ref. \cite{krolikowski2001modulational}, we can transform the linearized Eqs.  (\ref{coupled1}) and (\ref{coupled2}) into a set of ordinary differential equations in \(k\) space by utilizing the convolution theorem for the Fourier transformation. The resulting a set of coupled ODEs are
    \begin{subequations}
      \begin{align}
         \frac{\partial \hat{u}}{\partial \tau} & =\frac{\mu(\tau)}{2} k^2 \hat{v}, \label{FT_ODE1} \\
         \frac{\partial \hat{v}}{\partial \tau} & = \left(-\frac{\mu(\tau)}{2} k^2 + 2\rho\, p_0\, \hat{R}(k) \right) \hat{u}. \label{FT_ODE2}
       \end{align}  
    \end{subequations}

    The above equations (\ref{FT_ODE1}) and (\ref{FT_ODE2}) can be expressed in matrix form 
    \begin{equation}
        \frac{\partial}{\partial z} X = A X.
        \label{matrix_form}
    \end{equation}
Here the vector \( X \) and matrix \( A \) are defined as,
    \begin{equation}
       X = \begin{bmatrix}
             \hat{u} \\
             \hat{v}
           \end{bmatrix}, 
       A = \begin{bmatrix}
              0 & \frac{\mu(\tau) k^2}{2} \\
             -\frac{\mu(\tau) k^2}{2} + 2\rho p_0 \hat{R}(k) & 0
            \end{bmatrix}.
         \label{matrix_X_A}
    \end{equation}
The characteristic equation to find the eigenvalues $\lambda_{1,2}$ of the matrix $( A )$ is given by
 
    \begin{equation}
        \lambda^2 = -  k^2 p_0 ~ [ \alpha \mu(\tau)^2 k^2 - \mu(\tau)\rho\, \hat{R}(k) ].
        \label{eigenvalues2}
    \end{equation}
The above equation is the basis for our  MI studies. Here we consider the Gaussian response function $R(x)=\frac{1}{\sqrt{\pi} \sigma} \exp \left(-\frac{x^2}{\sigma^2}\right)$, which is  real and symmetric function. Therefore, the Fourier transform of this Gaussian response function is given by  
$\hat{R}(k) = \exp(-\sigma^2 k^2 / 4)$. Therefore, the growth rate of the perturbation is identified as
    \begin{multline}
        |\Re(\lambda)| = \\
        \Re \left( \sqrt{\mu(\tau) k^2 \rho p_0 \exp(-\sigma^2 k^2 / 4) - \frac{\mu(\tau)^2 k^4}{4}} \right).
        \label{MI_growth_rate}
    \end{multline}

\begin{figure*}[hbt!]
       \centering
         \includegraphics[scale=0.7]{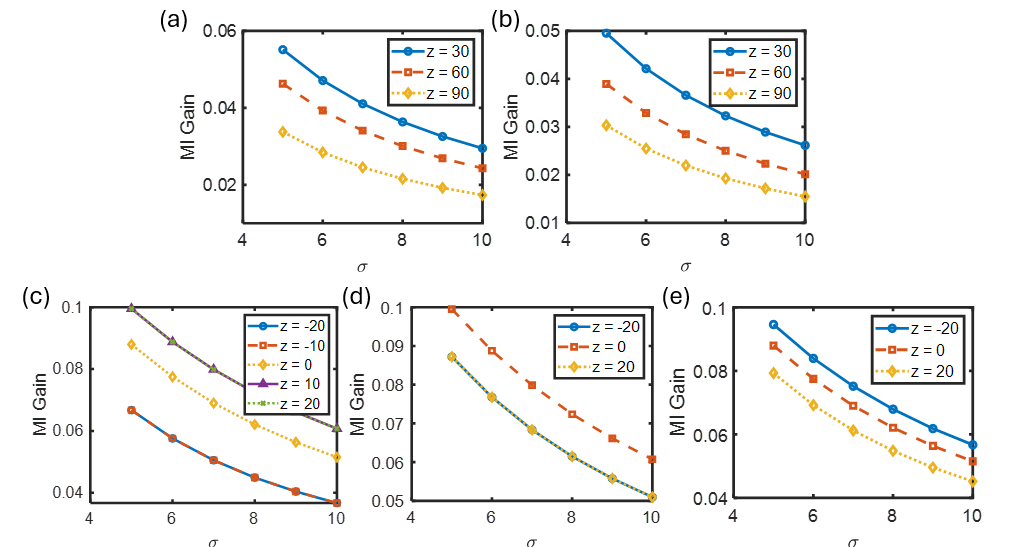}
         \caption{MI gain plotted at different propagation distances for varying values of the nonlocality parameter $\sigma$: (a) linearly decreasing diffraction, (b) exponentially decreasing diffraction, (c) step-like diffraction, (d) barrier-type diffraction, and (e) periodic diffraction profiles.}
       \label{MI_growth}
    \end{figure*}

\begin{figure*}[hbt!]
        \centering
            \includegraphics[scale=0.6]{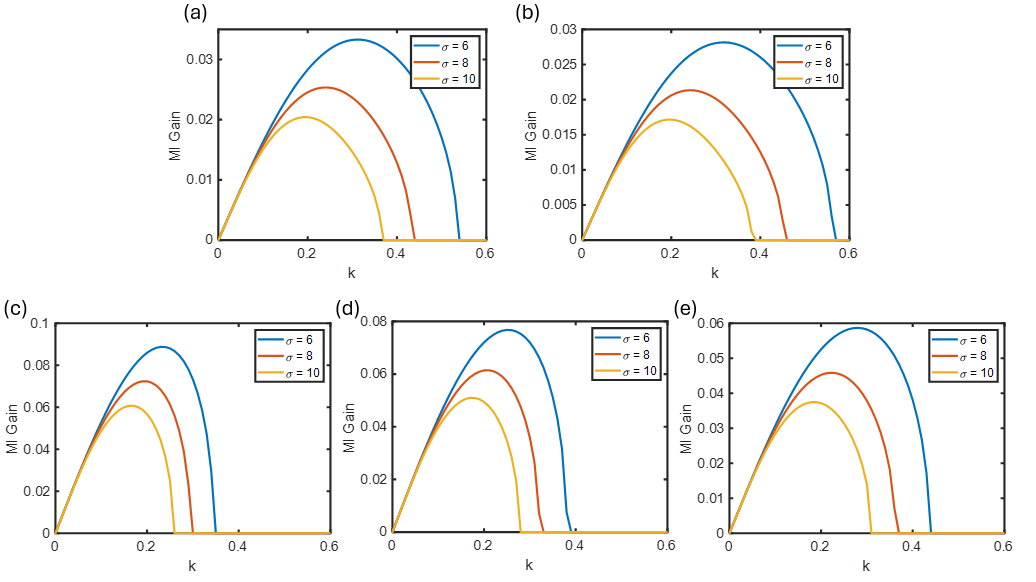}
             \caption{MI bandwidth for different values of the nonlocality parameter [\(\sigma = 6\) (blue), \(\sigma = 8\) (red), \(\sigma = 10\) (yellow)]: (a) linearly decreasing, (b) exponentially decreasing, (c) step-like, (d) barrier-type, and (e) periodic diffraction profiles.}
        \label{MIbandwidth}
    \end{figure*}

    In Fig.~\ref{MI_growth}, we present the MI gain corresponding to the five distinct forms of diffraction modulation, with $\rho>0$, by varying the length of the response function $\sigma$, which characterizes the strength of nonlocality, at different propagation distances. Figure~\ref{MI_growth}(a) depicts how the MI gain varies with the nonlocality parameter $\sigma$ for the linear diffraction case at three different longitudinal positions. Specifically, the MI gain is plotted for $z = 30$ (blue curve), $z = 60$ (red curve), and $z = 90$ (yellow curve). From this figure, it can be observed that, for all propagation distances, the MI gain is suppressed as $\sigma$ increases. Consequently, the plane wave (\ref{planewave}) becomes more stable against MI as nonlocality increases.This result demonstrates that, under linear diffraction, the presence of strong nonlocal nonlinearity significantly suppresses the MI in the focusing nonlinearity regime. To confirm this behavior, we plot the MI gain as a function of the perturbation wave number in Fig.~\ref{MIbandwidth} for various values of the nonlocality parameter $\sigma$. As seen in Fig.~\ref{MIbandwidth}(a), MI is restricted to a narrow wave number band for strongly nonlocal focusing nonlinearity ($\sigma=10$), whereas for $\sigma=6$ the nonlocal nonlinearity exhibits a broader MI band, thereby increasing the possibility of observing modulation instability. 

    As in the linear diffraction case, the MI gain behavior for exponentially decreasing diffraction modulation follows a similar trend, as illustrated in Figs.~\ref{MI_growth}(b) and \ref{MIbandwidth}(b). Nevertheless, the MI gain values are lower than in the linear case, while the MI gain bandwidth increases slightly. Figure \ref{MI_growth}(c) depicts the MI gain versus the nonlocality parameter $\sigma$ for the step-like diffraction profile at five different propagation distances. Unlike the previous cases, the MI gain curves in the positive $z$-direction overlap, indicating that the gain remains nearly constant along the positive $z$-direction. Similarly, in the negative $z$-direction, the MI gain decreases with increasing $\sigma$ at all distances. This demonstrates that, for the step-like profile, the MI dynamics are nearly symmetric with respect to the propagation direction. However, as shown in Fig.~\ref{MIbandwidth}(c), the MI bandwidth is significantly reduced to low wave numbers, causing MI to occur less frequently in the step-like diffraction modulation. 

    In Fig.~\ref{MI_growth}(d), we present the results for the barrier-type diffraction profile. Unlike the step-like diffraction case, the MI gain at $z = 0$ is higher than at other propagation distances. Interestingly, the curves corresponding to $z = -20$ and $z = 20$ overlap, indicating symmetric behavior in both the positive and negative propagation directions. Thus, the barrier-type diffraction modulation enhances the MI gain at the center while preserving this symmetry. The corresponding MI bandwidth is displayed in Fig.~\ref{MIbandwidth}(d), where the MI region is slightly broadened compared to the narrower region observed for the step-like diffraction modulation. Therefore, MI is more likely to occur than in the step-like diffraction profile case.

    Finally, Fig.~\ref{MI_growth}(e) shows the results for the periodic diffraction profile. In this case, a clear distance-dependent behavior is observed. That is, the MI gain gradually decreases with increasing $z$ in both the positive and negative propagation directions. Unlike the step-like and barrier-type diffraction modulation cases, where the MI gain curves overlap and suppression is strong at large $z$, the sinusoidal profile exhibits a moderate but consistent reduction of MI gain with distance. This indicates that the periodic diffraction profile supports a broader MI bandwidth, leading to a higher possibility of plane-wave instability, as we demonstrated in Fig. \ref{MIbandwidth}(e).

   \section{Conclusion}
    This work presents a comprehensive study on the dynamics of optical pulse propagation in a strongly nonlocal nonlinear medium with spatially varying diffraction profiles. By employing a combination of analytical and numerical techniques, including the LVM and RK4 method, we have investigated how linear, exponential, step-like, barrier-type, and periodic diffraction modulations influence the beam characteristics. The numerical simulations based on the split-step Fourier method show excellent agreement with the results obtained from the variational analysis and the RK4 method. In all cases, we found that the strongly nonlocal medium with variable diffraction supports diffraction-managed breather solitons.

   In addition, we have investigated the influence of spatially varying diffraction profiles on the modulation instability dynamics in a strongly nonlocal nonlinear medium. By analysing the MI gain spectra for the considered different diffraction profiles, we observed that the diffraction landscape plays a crucial role in shaping both the strength and bandwidth of instability. Specifically, the exponentially decreasing diffraction profile supports the broadest MI bandwidth, whereas the step-like and barrier-type diffraction profiles exhibit the highest peak gain and a narrower window for MI occurrence. 
   The periodic and linearly decreasing diffraction modulations demonstrate intermediate MI growth behavior. The results obtained in this paper confirm that tailoring the diffraction profile of the medium provides an effective way to control the propagation dynamics and to observe the formation of diffraction-managed breathing solitons. The diffraction engineering offers an additional degree of freedom for stabilising beam propagation, with potential applications in the design of all-optical systems, beam shaping, and nonlinear light control.

    \section{Acknowledgment}
    KN acknowledges the Department of Science \& Technology (DST) for support through the Core Research Grant (CRG/2023/008068), CEFIPRA/IFCPAR (IFC/7148/2023), and UKIERI-SPARC (3673)  for financial support through research projects. Mamatha K L gratefully acknowledges CSIR for the financial support (File No: 09/1001(13501)/2022-EMR-I). SS acknowledges DST-ANRF, India, for the financial support received from the National Science Chair (NSC/2020/000029) project of Prof. M. Lakshmanan.
    
    \section{Appendix}
   The propagation of an optical pulse in a nonlinear medium is described using the standard paraxial wave equation, which can be deduced by considering the slowly varying approximation, for the evolution of the slowly varying complex envelope \(\psi(x,z)\):
    \begin{equation}
    i\,\frac{\partial \psi}{\partial z}
    + \frac{1}{2k n_0}\,\frac{\partial^2 \psi}{\partial x^2}
    + \rho\,\Delta n\,\psi = 0 .
    \end{equation}
    Here, \(k\) is wave number in the medium, \(n_0\) is the refractive index of the medium, \(\rho\) denotes the nonlinearity coefficient, and \(\Delta n\) represents the intensity induced nonlinear refractive-index. In the present system (\ref{NNLSE}), we assumed that the refractive index is not uniform but varies along the propagation direction due to the inhomogeneity of the nonlocal medium. Therefore, it is appropriate to replace the constant refractive index ($n_0$) by a longitudinally varying one, i.e., \(n_0 \rightarrow n(z)\).
    In a nonlocal nonlinear medium, the refractive-index change depends on the surrounding intensity distribution and is given by
    \begin{equation}
    \Delta n(x,z) = \int_{-\infty}^{\infty} R(x-\xi)\,I(\xi,z)\,d\xi ,
    \end{equation}
    where \(R(x)\) is the nonlocal response function and \(I(\xi,z)=|\psi(\xi,z)|^2\) is the optical intensity.
    With these considerations, the paraxial wave equation now takes the form
    \begin{equation}
    i\,\frac{\partial \psi}{\partial z}
    + \frac{1}{2k n(z)}\,\frac{\partial^2 \psi}{\partial x^2}
    + \rho\,\Delta n(x,z)\,\psi = 0, 
    \end{equation}
    The above equation can be simply written as \begin{equation}
    i\,\frac{\partial \psi}{\partial z}
    + \frac{\mu(z)}{2}\,\frac{\partial^2 \psi}{\partial x^2}
    + \rho\,\Delta n(x,z)\,\psi = 0, \label{29}
    \end{equation}
   where  the longitudinally varying diffraction coefficient $\mu(z)$ is defined as
    \begin{equation}
    \mu(z) = \frac{1}{k n(z)}.
    \end{equation}
We wish to point out here that the above form of the vc-NNLSE in Eq. (\ref{29}) can be realized from a one-dimensional nematic liquid crystal model by following the theory developed in \cite{rasmussen2005theory} and by appropriately engineering the optical axis of the NLC.

\bibliographystyle{elsarticle-num} 
\biboptions{numbers,sort&compress}

\begin{thebibliography}{10}
\expandafter\ifx\csname url\endcsname\relax
  \def\url#1{\texttt{#1}}\fi
\expandafter\ifx\csname urlprefix\endcsname\relax\def\urlprefix{URL }\fi
\expandafter\ifx\csname href\endcsname\relax
  \def\href#1#2{#2} \def\path#1{#1}\fi

\bibitem{scott2003nonlinear}
A.~Scott, Nonlinear science: emergence and dynamics of coherent structures, Oxford University Press, 2003.

\bibitem{kivshar2003optical}
Y.~S. Kivshar, G.~P. Agrawal, Optical solitons: from fibers to photonic crystals, Academic press, 2003.

\bibitem{stegeman1999optical}
G.~I. Stegeman, M.~Segev, Optical spatial solitons and their interactions: Universality and diversity, Science 286 (1999) 1518.

\bibitem{snyder1997accessible}
A.~W. Snyder, D.~J. Mitchell, Accessible solitons, Science 276 (1997) 1538.

\bibitem{krolikowski2001modulational}
W.~Krolikowski, O.~Bang, J.~J. Rasmussen, J.~Wyller, Modulational instability in nonlocal nonlinear kerr media, Phys. Rev. E 64 (2001) 016612.

\bibitem{rasmussen2005theory}
P.~D. Rasmussen, O.~Bang, W.~Kr{\'o}likowski, Theory of nonlocal soliton interaction in nematic liquid crystals, Phys. Rev. E 72 (2005) 066611.

\bibitem{krolikowski2000solitons}
W.~Kr{\'o}likowski, O.~Bang, Solitons in nonlocal nonlinear media: Exact solutions, Phys. Rev. E 63 (2000) 016610.

\bibitem{rosanov2002internal}
N.~Rosanov, A.~Vladimirov, D.~Skryabin, W.~Firth, Internal oscillations of solitons in two-dimensional nls equation with nonlocal nonlinearity, Phys. Lett. A 293 (2002) 45.

\bibitem{guo2004large}
Q.~Guo, B.~Luo, F.~Yi, S.~Chi, Y.~Xie, Large phase shift of nonlocal optical spatial solitons, Phys. Rev E 69 (2004) 016602.

\bibitem{zhong2016spatiotemporal}
W.-P. Zhong, M.~R. Beli{\'c}, B.~A. Malomed, Y.~Zhang, T.~Huang, Spatiotemporal accessible solitons in fractional dimensions, Phys. Rev. E 94 (2016) 012216.

\bibitem{he2008crescent}
Y.~He, B.~A. Malomed, D.~Mihalache, H.~Wang, Crescent vortex solitons in strongly nonlocal nonlinear media, Phys. Rev. A 78 (2008) 023824.

\bibitem{ouyang2006perturbative}
S.~Ouyang, Q.~Guo, W.~Hu, Perturbative analysis of generally nonlocal spatial optical solitons, Phys. Rev. E 74 (2006) 036622.

\bibitem{10.1063/1.1713919}
J.~P. Gordon, R.~C.~C. Leite, R.~S. Moore, S.~P.~S. Porto, J.~R. Whinnery, Long‐transient effects in lasers with inserted liquid samples, Journal of Applied Physics 36 (1965) 3.

\bibitem{1074930}
S.~Akhmanov, D.~Krindach, A.~Migulin, A.~Sukhorukov, R.~Khokhlov, Thermal self-actions of laser beams, IEEE Journal of Quantum Electronics 4 (1968) 568.

\bibitem{PhysRevA.48.4583}
D.~Suter, T.~Blasberg, Stabilization of transverse solitary waves by a nonlocal response of the nonlinear medium, Phys. Rev. A 48 (1993) 4583.

\bibitem{segev1992photorefractive}
M.~Segev, B.~Crosignani, A.~Yariv, B.~Fischer, Spatial solitons in photorefractive media, Phys. Rev. Lett. 68 (1992) 923.

\bibitem{Gatz:98}
S.~Gatz, J.~Herrmann, Anisotropy, nonlocality, and space-charge field displacement in ($2+1$)-dimensional self-trapping in biased photorefractive crystals, Opt. Lett 23 (1998) 1176.

\bibitem{PhysRevLett.94.160401}
A.~Griesmaier, J.~Werner, S.~Hensler, J.~Stuhler, T.~Pfau, Bose-einstein condensation of chromium, Phys. Rev. Lett. 94 (2005) 160401.

\bibitem{MCLAUGHLIN199555}
D.~W. McLaughlin, D.~J. Muraki, M.~J. Shelley, {Xiao Wang}, A paraxial model for optical self-focussing in a nematic liquid crystal, Physica D: Nonlinear Phenomena 88 (1995) 55--81.

\bibitem{PhysRevE.66.036603}
M.~A. Karpierz, Solitary waves in liquid crystalline waveguides, Phys. Rev.E 66 (2002) 036603.

\bibitem{conti2003route}
C.~Conti, M.~Peccianti, G.~Assanto, Route to nonlocality and observation of accessible solitons, Phys. Rev. Lett. 91 (2003) 073901.

\bibitem{conti2004observation}
C.~Conti, M.~Peccianti, G.~Assanto, Observation of optical spatial solitons in a highly nonlocal medium, Phys. Rev. Lett. 92 (2004) 113902.

\bibitem{Hlpecseli1980}
H.~L. Pecseli, J.~J. Rasmussen, Nonlinear electron waves in strongly magnetized plasmas, Plasma Physics 22 (1980) 421.

\bibitem{mihalache2006three}
D.~Mihalache, D.~Mazilu, F.~Lederer, B.~Malomed, Y.~V. Kartashov, L.-C. Crasovan, L.~Torner, Three-dimensional spatiotemporal optical solitons in nonlocal nonlinear media, Phys. Rev. E 73 (2006) 025601.

\bibitem{he2008spinning}
Y.~He, B.~A. Malomed, D.~Mihalache, H.~Wang, Spinning bearing-shaped solitons in strongly nonlocal nonlinear media, Phys. Rev. A 77 (2008) 043826.

\bibitem{zhou2014propagation}
G.~Zhou, R.~Chen, G.~Ru, Propagation of an airy beam in a strongly nonlocal nonlinear media, Laser Phys. Lett. 11 (2014) 105001.

\bibitem{peccianti2002nonlocal}
M.~Peccianti, K.~A. Brzd{\k{a}}kiewicz, G.~Assanto, Nonlocal spatial soliton interactions in nematic liquid crystals, Opt. Lett. 27 (2002) 1460.

\bibitem{hu2006nonlocality}
W.~Hu, T.~Zhang, Q.~Guo, L.~Xuan, S.~Lan, Nonlocality-controlled interaction of spatial solitons in nematic liquid crystals, Appl. Phys. Lett. 89 (2006).

\bibitem{chen2015interaction}
W.~Chen, M.~Shen, Q.~Kong, Q.~Wang, The interaction of dark solitons with competing nonlocal cubic nonlinearities, J. Opt 44 (2015) 271--280.

\bibitem{dreischuh2006observation}
A.~Dreischuh, D.~N. Neshev, D.~E. Petersen, O.~Bang, W.~Krolikowski, Observation of attraction between dark solitons, Phys. Rev. Lett. 96 (2006) 043901.

\bibitem{rotschild2006long}
C.~Rotschild, B.~Alfassi, O.~Cohen, M.~Segev, Long-range interactions between optical solitons, Nat. Physics 2 (2006) 769.

\bibitem{buccoliero2009quasi}
D.~Buccoliero, A.~S. Desyatnikov, Quasi-periodic transformations of nonlocal spatial solitons, Opt. Express 17 (2009) 9608.

\bibitem{dai2015propagation}
Z.~Dai, Z.~Yang, S.~Zhang, Z.~Pang, Propagation of anomalous vortex beams in strongly nonlocal nonlinear media, Opt. Commun. 350 (2015) 19--27.

\bibitem{PhysRevE.63.016610}
W.~Kr\'olikowski, O.~Bang, Solitons in nonlocal nonlinear media: Exact solutions, Phys. Rev. E 63 (2000) 016610.

\bibitem{PhysRevA.82.063829}
E.~N. Tsoy, Solitons in weakly nonlocal media with cubic-quintic nonlinearity, Phys. Rev. A 82 (2010) 063829.

\bibitem{kong2010analytical}
Q.~Kong, Q.~Wang, O.~Bang, W.~Krolikowski, Analytical theory for the dark-soliton interaction in nonlocal nonlinear materials with an arbitrary degree of nonlocality, Phys. Rev. A 82 (2010) 013826.

\bibitem{Tur85}
S.~K. Turitsyn, Spatial dispersion of nonlinearity and stability of multidimensional solitons, Theor. Math. Phys. 64 (1985) 797, translation from Teoret. Mat. Fiz., 1985, 64(2), 226--232.

\bibitem{PhysRevE.74.036622}
S.~Ouyang, Q.~Guo, W.~Hu, Perturbative analysis of generally nonlocal spatial optical solitons, Phys. Rev. E 74 (2006) 036622.

\bibitem{assanto2020spin}
G.~Assanto, N.~F. Smyth, Spin-optical solitons in liquid crystals, Phys. Rev. A 102 (2020) 033501.

\bibitem{sala2017bending}
F.~A. Sala, N.~F. Smyth, U.~A. Laudyn, M.~A. Karpierz, A.~A. Minzoni, G.~Assanto, Bending reorientational solitons with modulated alignment, JOSA B 34 (2017) 2459--2466.

\bibitem{huang2020diffraction}
S.-Y. Huang, B.-Y. Huang, C.-C. Kang, C.-T. Kuo, Diffraction and polarization properties of electrically--tunable nematic liquid crystal grating, Polymers 12 (2020) 1929.

\bibitem{peccianti2000electrically}
M.~Peccianti, A.~De~Rossi, G.~Assanto, A.~De~Luca, C.~Umeton, I.~Khoo, Electrically assisted self-confinement and waveguiding in planar nematic liquid crystal cells, Appl. Phys. Lett. 77 (2000) 7--9.

\bibitem{peccianti2004routing}
M.~Peccianti, C.~Conti, G.~Assanto, A.~De~Luca, C.~Umeton, Routing of anisotropic spatial solitons and modulational instability in liquid crystals, Nature 432 (2004) 733--737.

\bibitem{peccianti2005interplay}
M.~Peccianti, C.~Conti, G.~Assanto, Interplay between nonlocality and nonlinearity in nematic liquid crystals, Opt. Lett. 30 (2005) 415--417.

\bibitem{assanto2003nematicons}
G.~Assanto, M.~Peccianti, C.~Conti, Nematicons: optical spatial solitons in nematic liquid crystals, Optics and photonics news 14 (2003) 44--48.

\bibitem{calisto2023direction}
E.~Calisto, G.~Assanto, Direction-dependent optical solitary waves in nematic liquid crystals, Phys. Rev. A 108 (2023) 043509.

\bibitem{sabari2015dynamical}
S.~Sabari, C.~P. Jisha, K.~Porsezian, V.~A. Brazhnyi, Dynamical stability of dipolar {Bose--Einstein} condensates with temporal modulation of the s-wave scattering length, Phys. Rev. E 92 (2015) 032905.

\bibitem{cuevas2009solitons}
J.~Cuevas, B.~A. Malomed, P.~Kevrekidis, D.~Frantzeskakis, Solitons in quasi-one-dimensional {Bose--Einstein} condensates with competing dipolar and local interactions, Phys. Rev. A 79 (2009) 053608.

\bibitem{mishra2022energy}
M.~Mishra, S.~K. Kajala, M.~Sharma, S.~Konar, S.~Jana, Energy optimization of diffraction managed accessible solitons, JOSA B 39 (2022) 2804.

\bibitem{jia2012solitons}
J.~Jia, J.~Lin, Solitons in nonlocal nonlinear kerr media with exponential response function, Optics express 20 (2012) 7469--7479.

\bibitem{kong2013dark}
Q.~Kong, M.~Shen, Z.~Chen, Q.~Wang, R.-K. Lee, W.~Krolikowski, Dark solitons in nonlocal media with competing nonlinearities, Physical Review A—Atomic, Molecular, and Optical Physics 87 (2013) 063832.

\bibitem{liang2023solitons}
G.~Liang, F.~Shu, H.~Qiao, W.~Cheng, Solitons train in nonlocally nonlinear system with oscillatory responses, Chaos, Solitons \& Fractals 168 (2023) 113146.

\bibitem{Zhang:22}
D.~Zhang, L.~Zhong, Long-range interaction between solitons in strongly nonlocal nonlinear media with oscillatory response, J. Opt. Soc. Am. B 39 (2022) 3029--3034.

\bibitem{da1999analysis}
M.~Da~Silva, K.~Nobrega, A.~Sombra, Analysis of soliton switching in dispersion-decreasing fiber couplers, Opt. Communications 171 (1999) 351.

\bibitem{xiong2022analytical}
G.~Xiong, J.-R. He, K.~Wang, L.~Xue, Analytical light bullet solutions in diffraction-decreasing media with inhomogeneous parameters, Results in Physics 43 (2022) 106111.

\bibitem{malomed1997pulse}
B.~A. Malomed, Pulse propagation in a nonlinear optical fiber with periodically modulated dispersion: variational approach, Optics communications 136 (1997) 313--319.

\bibitem{anderson1983variational}
D.~Anderson, Variational approach to nonlinear pulse propagation in optical fibers, Phys. Rev. A 27 (1983) 3135.

\bibitem{anderson1999pereira}
D.~Anderson, F.~Cattani, M.~Lisak, On the pereira-stenflo solitons, Phys. Scripta 1999 (1999) 32.

\bibitem{freitas1998propagation}
D.~S. Freitas, J.~R. de~Oliveira, M.~A. de~Moura, Propagation of gaussian beams in a nonlinear medium, J. Phys. A: Math. Gen. 31 (1998) 1761--1769.

\bibitem{keshavarz2016gaussian}
A.~Keshavarz, M.~Naseri, Gaussian beam propagation in a kerr type metamaterial medium using abcd matrix method, Commun. Theor. Phys. 66 (2016) 231--236.

\bibitem{jovanoski1994propagation}
Z.~Jovanoski, R.~Sammut, Propagation of gaussian beams in a nonlinear saturable medium, Phys. Rev. E 50 (1994) 4087.

\bibitem{karlsson1992optical}
M.~Karlsson, Optical beams in saturable self-focusing media, Phys. Rev. A 46 (1992) 2726.

\bibitem{wang2018nonlocal}
Q.~Wang, J.~Li, X.~Gao, W.~Xie, Nonlocal logarithmic nonlinear optical soliton, Optik 172 (2018) 571.

\bibitem{tsoy2010solitons}
E.~N. Tsoy, Solitons in weakly nonlocal media with cubic-quintic nonlinearity, Phys. Rev. A 82 (2010) 063829.

\bibitem{mishra2022generation}
M.~Mishra, S.~K. Kajala, M.~Sharma, S.~Konar, S.~Jana, Generation, dynamics and bifurcation of high power soliton beams in cubic-quintic nonlocal nonlinear media, J. Opt 24 (2022) 055504.

\bibitem{agrawal2000nonlinear}
G.~P. Agrawal, Nonlinear fiber optics, in: Nonlinear Science at the Dawn of the 21st Century, Springer, 2000, p. 195.

\bibitem{krolikowski2004modulational}
W.~Krolikowski, O.~Bang, N.~I. Nikolov, D.~Neshev, J.~Wyller, J.~J. Rasmussen, D.~Edmundson, Modulational instability, solitons and beam propagation in spatially nonlocal nonlinearmedia, J. Opt. B: Quantum Semiclass. Opt. 6 (2004) S288.

\bibitem{tabi2022modulational}
C.~B. Tabi, H.~Tagwo, T.~C. Kofan{\'e}, Modulational instability in nonlinear saturable media with competing nonlocal nonlinearity, Phys. Rev. E 106 (2022) 054201.

\end{thebibliography}

\end{document}